\documentclass[11pt]{article}
\pdfoutput=1  

\usepackage{a4wide}
\usepackage{xspace}
\usepackage{multicol}
\usepackage{amsmath}
\usepackage{graphicx}
\usepackage{epsfig}
\usepackage{hyperref}
\usepackage{url}
\usepackage{amssymb}
\usepackage{color}

\newcommand{\be}{\begin{equation}}
\newcommand{\ee}{\end{equation}}
\newcommand{\ba}{\begin{eqnarray}}
\newcommand{\ea}{\end{eqnarray}}

\def\ud{\mathrm{d}}

\newcommand{\as}{\alpha_s}

\title{\textbf{Meaningful characterisation \\of perturbative theoretical uncertainties}}
\author{
  Matteo Cacciari$^{1,2}$ and Nicolas Houdeau$^{1}$ \\
\\
{\sl  \small $^1$LPTHE, UPMC Univ.~Paris 6 and CNRS UMR 7589, Paris, France}\\[2pt]
{\sl  \small $^2$Universit\'e Paris Diderot, Paris, France}\\[2pt]
}

\begin{document}

\maketitle

\vspace{-8.5cm}
 \begin{flushright}
   May 2011\\
 \end{flushright}
 \date
\vspace{8cm}

\begin{abstract}
We consider the problem of assigning a meaningful degree of belief to
uncertainty estimates of perturbative series. We analyse the assumptions which
are implicit in the conventional estimates made using renormalisation scale
variations. We then formulate a Bayesian model that, given equivalent initial
hypotheses, allows one to characterise a perturbative theoretical uncertainty in 
a rigorous way in terms of a credibility interval for the remainder of the 
series. We compare its outcome to the conventional uncertainty estimates in the 
simple case of the calculation of QCD corrections to the  $e^+e^- \to$ hadrons process. 
We find comparable results, but with important conceptual differences. This work represents 
a first step in the direction of a more comprehensive and rigorous handling of theoretical 
uncertainties in perturbative calculations used in high energy phenomenology.

\end{abstract}

\clearpage

\tableofcontents

\section{Introduction}

The Large Hadron Collider (LHC) has finally been fired up  and, no black hole
having swallowed the Earth, the race to collect data and analyse them has now started
in earnest. While the short term goal is to rediscover the Standard Model,
the long term one will of course be to find signals of ``new physics'', be it the
Higgs boson, supersymmetry, or something else, more exotic and possibly unexpected.
While it is everybody's hope that discoveries will announce themselves in the form of 
unambiguous signals, it is of course conceivable, and probably also unavoidable
initially, that they may rather present themselves cloaked under some subtle
data/theory discrepancy. If this is the case, a full control of the uncertainty of the
theoretical  predictions becomes naturally of paramount importance: when comparing an
experimental measurement to a theoretical calculation, we must be able to say if they
agree  or not, and with what degree of confidence we are making such statement. This
is impossible to achieve unless both the experiment and the theory are provided with a
meaningful (and commonly accepted) degree of uncertainty.

While most of what we discuss below can apply to any kind of theoretical prediction in
perturbation theory, we will specialize it to the context of  Quantum Chromodynamics
(QCD): many LHC processes and backgrounds pertain to the QCD realm and, due to the
relatively large size of the QCD coupling $\as$ and therefore the slower
perturbative convergence,  the issue of  theoretical accuracy is more
pressing. Theoretical predictions in QCD contain multiple ingredients, inputs that 
must be ultimately extracted from experimental data, 
like Parton Distribution Functions (PDFs) for hadronic collisions and the value of 
$\as$. 
In the past several years a lot of progress has been made on $\as$ and the PDFs.
The uncertainty with which we know the coupling is now quite small (see for instance \cite{Bethke:2009jm}). 
Moreover, several groups \cite{Martin:2009iq,Lai:2010vv,Ball:2010de,:2009wt,Alekhin:2009ni} have extracted PDF
sets with associated uncertainties of experimental origin, and provided frameworks to
properly propagate them to the observable one is calculating. Huge progress has also
been made in performing higher order perturbative  calculations for a large number of
phenomenologically interesting observables~\cite{Binoth:2010ra}, thereby potentially improving the accuracy
with which they are known.

One area where progress has instead arguably not been made is in an understanding of the
meaning of the residual theoretical uncertainty, given by unknown higher orders in
perturbation theory.  This uncertainty is usually  estimated by varying unphysical
momentum scales (we will denote them collectively by $\mu$) contained in the perturbative result, 
like the renormalisation and the
factorisation scales, around a central value $\mu_0$, usually taken to coincide with a
physical momentum scale $Q$ of the process. The method, the range in which to vary
the scales (typically $[\mu_0/2,2\mu_0]$), and their central value $\mu_0=Q$ are highly
conventional, but nevertheless quite commonly accepted. They allow the community to
efficiently exchange a conventional uncertainty which can be easily compared
between different calculations. 

Among the shortcomings one may find in this procedure, the most glaring one is probably
that it does not allow one to estimate the {\sl degree of belief} (DoB) of the resulting
uncertainty band. 
By this we mean that it is not possible to associate a value, $68.3\%$, $95.5\%$ or $99.7\%$
for example, to our belief that the uncertainty band contains 
the true sum\footnote{\label{footnote:one} We
forego here the fact that QCD series are usually not convergent but simply asymptotic.
The onset of the asymptotic behaviour usually taking place at fairly high perturbative
orders, it normally  does not affect realistic phenomenological applications. In practice, 
the place of ``true sum of the series'' can be taken by the asymptotic value of the series calculated with an appropriate prescription, or even by some more refined higher order result, though we will keep (mis)using the term ``true result'' to mean the desired result beyond what has been really calculated.} of the series. This lack of a proper characterisation of the perturbative theoretical uncertainty also means that procedures to combine it with other sources of uncertainties (e.g. the value of the coupling and the PDFs) are at best ambiguous or controversial, as exemplified by a recent discussion \cite{Baglio:2010um} about the proper way to estimate the total uncertainty of the prediction for the Higgs production cross section at hadron colliders.
All this
makes it potentially impossible to fully and rigorously assess our degree of belief that an experimental result may agree (or not) with theory, making betting (or, in order  to cover both sides, offering odds) on new physics having been discovered or not an altogether unscientific -- and potentially risky -- proposition.

The purpose of this paper is precisely to try to make such a bet potentially safe, consistently with the {\sl coherent bet} idea of de Finetti~\cite{deFinetti}. To achieve this we construct a model that leads to a well defined credibility measure for a perturbative theoretical uncertainty, so that the degree of belief of a given interval can be explicitly calculated.
In section~\ref{sect:def} we first review the commonly used theoretical uncertainty estimation via unphysical scales variations, and subsequently  proceed to define the Bayesian model from which we then extract our credibility distributions.
Section~\ref{sect:compVarMu} compares the results of our credibility-based model with those of the conventional method, allowing one to assign a degree of belief to the uncertainty bands given by the latter. Some results for the $e^+e^- \to$ hadrons process are given in section~\ref{sect:results}, to better illustrate the model with a realistic example. Section~\ref{sect:modifModel} discusses some of the hypotheses that were made in building the model, and Section~\ref{sect:limModel} extends it to the case of partial knowledge of higher order coefficients.

Before closing this introduction we wish to stress the following point: we are {\sl not} trying to improve our knowledge of a perturbative prediction by adding physical information or even just speculations about its form, or by (improbably) seeking physical content inside the mathematical formalism: the only information that enters the result is what has been explicitly calculated, i.e.  the known coefficients of a perturbative series. To this information we add hypotheses meant to formalize assumptions that are often implicitly made when estimating theoretical uncertainty using scale variations, and we use the framework of Bayesian probability (see section~\ref{sect:confidence}) for computing from them and from the available information the degree of belief of given uncertainty intervals. The hypotheses need not even be strictly true (or people may disagree about them), but once they are made the path to the calculation of the degree of belief values is a rigorous one.


\section{Theoretical uncertainty estimates}
\label{sect:def}

For definiteness, consider the perturbative calculation for  the cross section of a process taking place at a hard scale $Q$ (see footnote~\ref{footnote:one} for a comment about the asymptotic nature of QCD series):
\be
\sigma(Q) = \sum_{n=0}^\infty c_n(Q,\mu_R) \as^n(\mu_R) \, ,
\ee
where $\mu_R$ is the renormalisation scale (which we shall in the following simply denote by $\mu$), and the coupling $\as(\mu)$ evolves according to 
\be
\frac{\ud\as}{\ud\ln\mu^2} = \beta(\as) = -\as^2\sum_{n=0}^\infty \beta_n \as^n \; .
\ee
A concrete example would of course be the production of hadrons in $e^+e^-$ collisions. When no dependence is given explicitly, the coefficients and the coupling will be considered to be evaluated at a renormalisation scale $\mu= Q$:
\be
\sigma(Q) = \sum_{n=0}^\infty c_n(Q,Q) \as^n(Q) \equiv \sum_{n=0}^\infty c_n \as^n\; .
\ee
Given $c_n \equiv c_n(Q,Q)$ independent of $\mu$, one can always reinstate the full $\mu$ dependence and determine $c_n(Q,\mu)$ using 
\be\label{eq:defCn}
c_n(Q,\mu) = \sum_{l=0}^{n-1}c_{n,l}\left(\ln\frac{\mu^2}{Q^2}\right)^l
\ee
where $c_{n,0} = c_n$ and 
\be\label{eq:defCnl}
c_{n,l}=\frac{1}{l}\sum_{j=0}^{n-1}j\beta_{n-1-j}c_{j,l-1}
\ee
(see appendix~\ref{app:coefs} for a derivation). Note that this last equation uses all the coefficients $c_j$ with $j < n$.

We will also denote by
\be\label{eq:defSk}
\sigma_{k}(Q,\mu) \equiv \sum_{n=0}^k c_n(Q,\mu) \as^n(\mu)
\ee
(or, for short, $\sigma_{k} \equiv \sum_{n=0}^k c_n \as^n$ for $\mu=Q$) the partial sum up to the last calculated perturbative order $k$, and by 
\be\label{eq:defDk}
\Delta_k \equiv \sum_{n=k+1}^\infty c_n \as^n. 
\ee
the remainder. 

\subsection{Conventional theoretical uncertainty estimate}
\label{sect:conv}

The explicit $\mu$ dependence of $\sigma_k(Q,\mu)$ in eq.~(\ref{eq:defSk}) serves as a reminder that, when truncated to a finite order, a perturbative calculation retains a higher-order dependence on the scale $\mu$. This dependence is generally 
exploited to estimate its ``uncertainty''\footnote{Of course it is not so much $\sigma_k$ which is ``uncertain'', in that it is perfectly well determined by the knowledge of its coefficients and parameters, but rather the true value of the series and therefore to what extent $\sigma_k$ describes it.}, i.e. the presumed value of $\Delta_k$.  In order to do so, one typically quotes an uncertainty interval $[\sigma_k^-,\sigma_k^+]$ around $\sigma_k$ (but not necessarily centred on it). The specific choices for $\sigma_k^\pm$ can vary. Possible options are:
\begin{enumerate}
\item
\be
\label{eq:sigmak1}
\sigma_k^- = \min\{\sigma_k(Q,Q/2),\sigma_k(Q,2Q)\}\qquad\qquad\sigma_k^+ = \max\{\sigma_k(Q,Q/2),\sigma_k(Q,2Q)\}
\ee
\item
\be
\label{eq:sigmak2}
\sigma_k^- = \min_{\mu\in[Q/2,2Q]}\{\sigma_k(Q,\mu)\}\qquad\qquad \sigma_k^+ = \max_{\mu\in[Q/2,2Q]}\{\sigma_k(Q,\mu)\}
\ee
\item
\be
\label{eq:conv}
\sigma_k^\pm = \sigma_k \pm \frac{\delta_k}{2}
\ee
where 
\be
\delta_k \equiv |\sigma_k(Q,2Q)-\sigma_k(Q,Q/2)|
\ee
\item
Same as eq.~(\ref{eq:conv}), but with
\be
\delta_k \equiv \max_{\mu\in[Q/2,2Q]}\{\sigma_k(Q,\mu)\}-\min_{\mu\in[Q/2,2Q]}\{\sigma_k(Q,\mu)\}
\ee
\end{enumerate}
In the last two cases  the interval is centred on $\sigma_k(Q,Q)$, whereas in the first two it is not necessarily so. Note also that the choice of varying the scale $\mu$ within a factor of two around the physical scale $Q$, i.e. in the range $[Q/2,2Q]$,  is fully conventional.

A priori there is no reason  why the interval $[\sigma_k^-,\sigma_k^+]$ should represent a sensible estimate of the remainder  $\Delta_k$ of the series since, from a purely mathematical point of view, $\delta_k$ (or $\sigma_k$) does not contain any information about $\Delta_k$:  $\sigma_k(Q,\mu)$ and $\delta_k$ are functions of the $c_n$ for $n\leq k$, while $\Delta_k$ is a function of the $c_n$ for $n>k$.
However, the reason why this can instead often work in practice is that, under certain circumstances, 
the size of $\delta_k$  can be similar to the size of $\Delta_k$. One can indeed show that (see appendix~\ref{app:coefs})
\be
\label{eq:approxdelta1}
\delta_k \simeq \left|\frac{\ud\sigma_k}{\ud\ln\mu^2}\right|_{\mu=Q} [\ln (2Q)^2 -\ln (Q/2)^2] \simeq 3 k\beta_0 \as^{k+1}|c_k|
\ee
where the factor of 3 in the last term has been obtained by approximating the exact expression $4\ln 2$ (This factor would be replaced by $4\ln r$ if the scale $\mu$ were varied in the range $[Q/r,rQ]$).
The last equality above is obtained by making the assumption that all the coefficients in the series share the same magnitude and that $\as$ is reasonably small. Under these same hypotheses (and therefore $|c_{k+1}| \simeq |c_k|$), we can also write
\be
|\Delta_k| \simeq \as^{k+1} |c_{k+1}| \sim \delta_k \, .
\ee
Experience with perturbative  calculations in QCD has shown that theoretical uncertainty estimates like those of eq.~(\ref{eq:conv}) are quite successful in predicting the range in which a higher order result will fall. This can then be seen as an empirical validation of the assumption made above, i.e. that  $|c_{k+1}|$ is indeed often of the same magnitude as $|c_k|$.

The limitation of this conventional approach is that, even if the hypothesis $|c_{k+1}| \simeq |c_k|$ is correct and therefore $\delta_k$ correctly 
describes the size of the remainder of the series, there is no way of deciding how reliably it may do so.

\subsection{Credibility-based theoretical uncertainty estimate}
\label{sect:confidence}

In this paper we use the ``Bayesian probability'' (also called ``subjective probability'' or ``degree of belief'' or ``credibility'', see e.g. \cite{berger}), and distinguish it from the ``frequentist probability''. 
The two concepts share the same mathematical formalism, but
are nonetheless distinct. 
Bayesian probability is not 
linked to an infinite number of realizations of an experiment. It deals with a particular
question, which may or may not be about the result of one particular realization of a
given experiment, and the consequences of the information one considers about its
possible answer\footnote{One {\sl may} build his initial credibility distribution using information of frequentist origin: e.g. after throwing an unbiased six-sided dice a large number of times (and hence establishing a frequentist probability), one can come to believe (i.e. define a credibility measure)  that there is a one-in-six chance that a given number will show up in a subsequent throw (i.e. set the credibility measure  equal to the frequentist probability previously established).  However, information of non-frequentist origin can also be included in a credibility-based approach: if someone is told that the dice is likely crooked, they can then adjust their expectations (degree of belief) using this information, even before throwing the dice a single time.}. This information is not necessarily rigorous or ``true'' in any way, but its treatment, once translated mathematically into the so called ``priors'' and ``likelihoods'', is.
A distribution of frequentist probability (or, for instance, its variance) gives a measure of the {\sl reproducibility} of an experiment. 
Conversely, a credibility distribution conveys information about the {\sl uncertainty} of the answer to a question,
for instance the result of one particular realization of an experiment, prior to its execution.
The variables appearing in a frequentist probability distribution are commonly denoted as {\sl random variables}, since they take different values in different realizations of the experiment. We call instead {\sl uncertain variables} the ones in a credibility distribution, to better make the distinction with the former ones: their values are not random (each of them being a single number), but simply unknown.

Given a density
function $f$, the degree of belief (or ``credibility'') that the value of an uncertain variable $\eta$ belongs to the interval $[a,b]$ is then equal to
\be
\mathbb{C}(\eta\in[a,b])\equiv\int_a^b
f(\eta)\ud\eta\, .
\ee 
where the result is a number between zero and one. 
\footnote{Note that, while this may seem similar to the  ``confidence level intervals'' of frequentist statistical analyses, our intervals are to be understood strictly within a Bayesian framework (where they can also be called ``credible intervals'' at a given level), and should not be confused with the frequentist ones, which (see e.g.~\cite{dagostini}) do not in fact express a ``level of confidence''.}

In this paper we will always work within the concept of degree of belief as defined above, and will never use the frequentist probability. The latter is  not applicable to the case of a theoretical uncertainty, which is not amenable to a frequentist treatment (there is nothing one can ``repeat'') and is much more akin to a systematic uncertainty instead.

\subsubsection{The model}
\label{sect:model}

The goal of this paper is to establish a {\sl conditional density} $f(\Delta_k|c_0,\dots,c_k)$ for the value of remainder of the series $\Delta_k$ in eq.~(\ref{eq:defDk}), given the knowledge of the coefficients of the perturbative expansion up to order $k$, and study its behavior.  The reason for introducing a density function is that it contains much more information than a simple uncertainty band like the one established in eq.~(\ref{eq:conv}) in section~\ref{sect:conv}.

To achieve this we will create a generic credibility measure, applicable to any possible perturbative series, over the space of {\itshape a priori} unknown coefficients $c_0,c_1,\dots$. More precisely, we will create a density function $f(c_0,c_1,\dots)$, normalised such that
\be
\int f(c_0,c_1,\dots)\ \ud c_0\ \ud c_1\dots=1 
\ee
and whose parameters can be marginalised according to 
\begin{align}
\label{eq:defDensities}
f(c_0,\dots,c_{i-1},c_{i+1},\dots)&=\int f(c_0,\dots,c_{i-1},c_{i},c_{i+1},\dots)\ \ud c_i \, .\\
f(c_0,\dots,c_k)&=\int f(c_0,c_1,\dots)\ \ud c_{k+1}\ \ud c_{k+2}\dots \, .
\end{align}
In the case of one particular physical process, some coefficients will have been already computed up to order $k$: $c_0^{\text{true}},\dots,c_k^{\text{true}}$. The credibility measure for this particular process will be the inherited measure defined on the subspace corresponding to $c_0=c_0^{\text{true}}$, $c_1=c_1^{\text{true}}$, \dots, $c_k=c_k^{\text{true}}$. For brevity, we will use the notation $f(c_{k+1}|c_0,\dots,c_k)$ instead of $f(c_{k+1}|c_0=c_0^{\text{true}},\dots,c_k=c_k^{\text{true}})$. The density over the still unknown $c_{k+1}$ coefficient will then read, according to the standard conditional density rule,
\be
\label{eq:defCondDensities}
f(c_{k+1}|c_0,\dots,c_k)=\frac{f(c_{k+1},c_0,\dots,c_k)}{f(c_0,\dots,c_k)} \, .
\ee

To create this generic measure, we focus on the observation made at the end of section~\ref{sect:conv} 
(i.e. that the empirical success of conventional theoretical uncertainty estimates made using scale variations can be explained by the fact that successive perturbative coefficients have similar magnitudes)
and reverse it. {\sl We make the assumption that all the coefficients $c_n$ in a perturbative series share some sort of upper bound $\bar c>0$ to their absolute values, specific to the physical process studied.\footnote{This hypothesis is of course known to be violated in practice, for instance by the factorial growth of the coefficients in the presence of renormalons. However, knowing that such a factorial growth typically only kicks in at fairly large perturbative orders, we can safely assume that our hypothesis holds true for low perturbative orders, which are the ones which are calculated in practice. Another instance in which  our hypothesis is violated is in particular kinematical configurations (see e.g. \cite{Rubin:2010xp}), or when new production channels open up at some higher order.} The calculated coefficients will give an estimate of this $\bar c$, restricting the possible values for the unknown $c_n$.} The set of uncertain variables that define the space on which we will create our credibility measure is thus the set constituted of this parameter $\bar c$ and of all the {\itshape a priori} unknown coefficients.

The model rests on three hypotheses:
\begin{enumerate}
\item
{\bf Residual uncertainty}\\
We suppose that, if we happened to know beforehand the parameter $\bar c$, our residual  density for the value of an unknown coefficient $c_n$ would be in the form of a uniform distribution,
\be
\label{eq:cbar}
f(c_n|\bar c)=\frac{1}{2\bar c}\left\{
\begin{array}{cc}
1	&	\mbox{ if }	|c_n|\leq\bar c\\
0	&	\mbox{ if }	|c_n|>\bar c
\end{array}
\right.
\equiv \frac{1}{2\bar c} \chi_{|c_n|\leq\bar c} \, ,
\ee
where $\chi_A$ is the characteristic function of a set $A$.
We could (and probably should) use a density function that does not vanish anywhere, like a Gaussian distribution, but  the form (\ref{eq:cbar}) leads to simpler expressions, so we use it in the following to study the model analytically.\footnote{We have checked that using a Gaussian distribution of mean zero and standard deviation $\bar c$ does not significantly modify the general behaviour of the results shown in this paper.}
\item
{\bf Shared information and independence}\\
The parameter $\bar c$ 
models information that we consider to be shared by all coefficients, and we make it the only one. When $\bar c$ is known, the residual uncertainties on the values of two coefficients $c_n$ and $c_{n'}$ are then totally 
independent. In fact, we will suppose the coefficients to be mutually independent, so that
for a set of coefficients $\{c_i\}$ we have 
\be\label{eq:modIndep}
f(\{c_i, i\in I\}|\bar c)=\prod_{i\in I}f(c_i|\bar c) \, .
\ee
The value of $\bar c$ is the maximal information that the coefficients share. It corresponds to the maximal knowledge one could extract from the known coefficients $c_0,\dots,c_k$ in order 
to ``predict'' the possible values of unknown ones $c_n$, $n>k$.
\item
{\bf Hidden parameter}\\
The value of the $\bar c$ parameter is ``hidden'' in the knowledge of the $c_n$. As long as we have not  calculated any coefficient, we can only say that it is a positive real number, and that all values for its order of magnitude are {\itshape a priori} equally probable. In order to implement this in practice we define a  density for its logarithm as the limit of a uniform distribution between $|\ln\epsilon|$ and $-|\ln\epsilon|$ when a small parameter $\epsilon$ tends to zero:
\be
\label{eq:epsdep}
f_{\epsilon}(\ln\bar c)=\frac{1}{2|\!\ln\epsilon|}\ \chi_{|\ln\bar c|\leq|\ln\epsilon|}
\quad\Leftrightarrow\quad
f_{\epsilon}(\bar c)=\frac{1}{2|\!\ln\epsilon|}\frac{1}{\bar c}\ \chi_{\epsilon\leq\bar c\leq1/\epsilon}
\ee
We will perform calculations (both analytical and numerical) using this $\epsilon$-dependent density $f_\epsilon$ with $\epsilon\neq 0$, and the final result will then be the limit $\epsilon \to 0$. The vanishing of a density in this limit would mean that we do not have enough information to make any guess about the result. For example, $f_\epsilon(\ln \bar c)$ tends to a ``uniformly null'' density, meaning that that when no coefficients are known we have no information whatsoever about the possible value of $\ln \bar c$. 
\end{enumerate}

The three hypotheses (\ref{eq:cbar}), (\ref{eq:modIndep}) and (\ref{eq:epsdep}) define completely the credibility measure over the whole space of {\itshape a priori} uncertain variables $\{\bar c,c_0,c_1,\dots\}$. They then define every possible inherited measure on a subspace associated with a physical process whose first coefficients are known. 
Section~\ref{sect:modifModel} will revisit the choices made in building this model, for instance the choice of a uniform distribution for $\ln\bar c$ rather than for $\bar c$ (eq.~(\ref{eq:epsdep})), and study some alternatives and their consequences on the results.

The following subsections are dedicated to deriving from these hypotheses the densities $f(\bar c|c_0,\dots,c_k)$, $f(c_n|c_0,\dots,c_k)$ for $n>k$, and finally the  residual theoretical uncertainty of a perturbative prediction calculated up to order $k$,  $f(\Delta_k|c_0,\dots,c_k)$.

\subsubsection{Conditional densities $f(\bar c|c_0,\dots,c_k)$ and $f(c_{n}|c_0,\dots,c_k)$, $n>k$}

Using the three hypotheses eqs. (\ref{eq:cbar}), (\ref{eq:modIndep}) and (\ref{eq:epsdep}) and the properties of conditional densities, one can show that
\be
\label{eq:cbarKnowCk1}
f(\bar c|c_0,\dots,c_k)=(k+1)\ \frac{(\max(|c_0|,\dots,|c_k|))^{k+1}}{\bar c^{k+2}}\ \chi_{\bar c>\max(|c_0|,\dots,|c_k|)}
\ee
and
\be
\label{eq:cnKnowCk1}
f(c_n|c_0,\dots,c_k) =
\frac{1}{2}\frac{k+1}{k+2}\ \frac{(\max(|c_0|,\dots,|c_k|))^{k+1}}{(\max(|c_n|,|c_0|,\dots,|c_k|))^{k+2}} \, .
\ee

Let us derive for instance the second of these results. The conditional density for a generic (uncalculated) coefficient $c_n$, $n>k$, is by definition (see eq.~(\ref{eq:defCondDensities}))
\be
\label{eq:bayes}
f_\epsilon(c_n|c_0,\dots,c_k) = \frac{f_\epsilon(c_0,\dots,c_k,c_n)}{f_\epsilon(c_0,\dots,c_k)} \, .
\ee
As stated in the previous subsection, we perform all calculations with $\epsilon\neq 0$ and we take the $\epsilon\to 0$ limit at the end. 
From eq.~(\ref{eq:defDensities}) and the property of conditional densities, similar to eq.~(\ref{eq:defCondDensities}), we have
\begin{align}
\label{eq:cond}
f_\epsilon(c_0,\dots,c_k) 
	&=\int f_\epsilon(c_0,\dots,c_k,\bar c)\ \ud\bar c\nonumber \\
	&=\int f_\epsilon(c_0,\dots,c_k|\bar c) f_\epsilon(\bar c)\ \ud\bar c \, .\\
\intertext{
Using the factorisation property (\ref{eq:modIndep}) and the definitions~(\ref{eq:cbar}) and~(\ref{eq:epsdep}) we get
}
f_\epsilon(c_0,\dots,c_k) 	
        &=\int \left(\prod_{i=0}^k f(c_i|\bar c)\right)f_\epsilon(\bar c)\ \ud\bar c \nonumber\\
	&=\int \left(\prod_{i=0}^k \frac{1}{2\bar c} \chi_{|c_i|\leq\bar c} \right)
		\frac{1}{2|\!\ln\epsilon|}\frac{1}{\bar c}\ \chi_{\epsilon\leq\bar c\leq1/\epsilon}
		\ \ud\bar c \nonumber\\
	&=\frac{1}{2^{k+2}} \frac{1}{|\!\ln\epsilon|}\int^{1/\epsilon}_{\max(|c_0|,\dots,|c_k|,\epsilon)} \frac{1}{\bar c^{k+2}}\ \ud\bar c \, .\label{eq:c0ckeps}\\
\intertext{
A similar result holds for $f_\epsilon(c_0,\dots,c_k,c_n)$:
}
f_\epsilon(c_0,\dots,c_k,c_n) 
	&= \frac{1}{2^{k+3}} \frac{1}{|\!\ln\epsilon|}\int^{1/\epsilon}_{\max(|c_n|,|c_0|,\dots,|c_k|,\epsilon)} \frac{1}{\bar c^{k+3}}\ \ud\bar c \, .
\end{align}
We therefore can write, using eq.~(\ref{eq:bayes}),
\be
\label{eq:cnKnowCk}
f(c_n|c_0,\dots,c_k) =\lim_{\epsilon\to 0} \frac{f_\epsilon(c_0,\dots,c_k,c_n)}{f_\epsilon(c_0,\dots,c_k)} = 
\frac{1}{2}\frac{k+1}{k+2}\ \frac{(\max(|c_0|,\dots,|c_k|))^{k+1}}{(\max(|c_n|,|c_0|,\dots,|c_k|))^{k+2}} \, .
\ee
Note that in this equation the value $k+1$ represents the total {\sl number} of known perturbative coefficients $c_0,\dots,c_k$ used to estimate $c_n$ with $n > k$, rather than simply one unit above the last calculated perturbative order $k$. Similarly, $k+2$ is this total number plus one. If a series starts at a non-zero order $\as^l$, its last known perturbative order $k$ plus one will not give anymore the number of known coefficients. We will detail in section~\ref{sect:non-zero-start} the modifications to be made in this and in the following equations to account for such a case.

The derivation of eq.~(\ref{eq:cnKnowCk}) is given in some more detail in appendix~\ref{app:cnKnowCk1}. The full derivation of eq.~(\ref{eq:cbarKnowCk1}) is given in appendix~\ref{app:cbarKnowCk1}.
From now on we will collect the derivations of the densities and uncertainty intervals in appendix~\ref{app:derivDens}, since we do not wish to focus on the technicalities of the derivation of the conditional densities but rather on their behavior.

\begin{figure}[tp]
\begin{center}
\includegraphics[height=5.5cm]{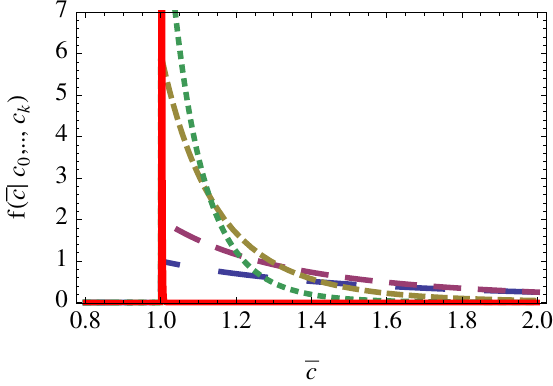}
~~
\includegraphics[height=5.5cm]{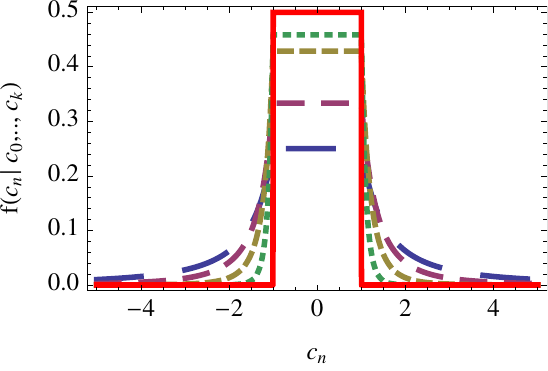}
\caption{Densities $f(\bar c|c_0,\dots,c_k)$ and $f(c_n|c_0,\dots,c_k)$ in the case $\bar c_{(k)}=1$ for $k=0$, $1$, $5$, $10$ and $10^3$ from the largest dashing to the solid curve.}
\label{fig:dens}
\end{center}
\end{figure}

Defining $\bar c_{(k)} \equiv \max(|c_0|,\dots,|c_k|)$ we can rewrite the densities (\ref{eq:cbarKnowCk1}) and (\ref{eq:cnKnowCk1}) as
\be
\label{eq:cbarKnowCk}
f(\bar c|c_0,\dots,c_k)=(k+1)\ \frac{{\bar c_{(k)}}^{k+1}}{\bar c^{k+2}}\ \chi_{\bar c> \bar c_{(k)}}
\ee
and
\be
\label{eq:fcn}
f(c_n|c_0,\dots,c_k)=\left(\frac{k+1}{k+2}\right)\frac{1}{2\bar c_{(k)}}\left\{
\begin{array}{cc}
1	&	\mbox{ if }	|c_n|\leq \bar c_{(k)}\\
\frac{1}{(|c_n|/\bar c_{(k)})^{k+2}}	&	\mbox{ if }	|c_n|>\bar c_{(k)}
\end{array}
\right. \, .
\ee
Figure \ref{fig:dens} shows the two distributions $f(c_n|c_0,\dots,c_k)$ and $f(\bar c|c_0,\dots,c_k)$, plotted as functions of $c_n$ and $\bar c$ respectively, for different values of $k$, assuming that $\bar c_{(k)}$ remains equal to one. $\bar c_{(k)}$ acts as an estimate of $\bar c$, and the density $f(\bar c|c_0,\dots,c_k)$ can be seen to tend to a Dirac distribution concentrated at $\bar c = \bar c_{(k)}$ when $k$ goes to infinity. The more coefficients are known, the more precisely $\bar c$ is estimated. 
In the same $k\to \infty$ limit the density over the unknown $c_n$ tends to $f(c_n|\bar c=\bar c_{(k)})$ as given in eq.~(\ref{eq:cbar}), the distribution that corresponds by construction to the remaining uncertainty when the whole of the hidden information simulated by $\bar c$ is known. For a finite value of $k$, the density is always wider than this limit: the uncertainty about unknown coefficients $c_n$ is  larger when one  knows the values of only a few coefficients $c_0,\dots,c_k$ than when one posses the full information about the value of $\bar c$. 

\subsubsection{Conditional  density $f(\Delta_k|c_0,\dots,c_k)$}
\label{sect:deltak}

The remainder $\Delta_k$ of the perturbative series depends on the values of all the unknown coefficients $c_{k+1},c_{k+2},\dots$. Its density can be written as
\be
\label{eq:DeltaKnowCkExact}
f(\Delta_k|c_0,\dots,c_k)=\int \left[\delta(\Delta_k-\sum_{n=k+1}^{\infty}\as^n c_n)\right]f(c_{k+1},c_{k+2},\dots|c_0,\dots,c_k)\ \ud c_{k+1} \ud c_{k+2} \dots
\ee
This expression is too complicated to be handled analytically, even in the case of the simple choice of density in eq.~(\ref{eq:cbar}) for the coefficients. However, making the approximation 
\be\label{eq:approxDelta}
\Delta_k\simeq\as^{k+1} c_{k+1} \, ,
\ee
and using eq.~(\ref{eq:fcn}) for $f(c_n|c_0,\dots,c_k)$ with $n=k+1$, we obtain 
\be
\label{eq:DeltaKnowCkExpression}
f(\Delta_k|c_0,\dots,c_k)\simeq\left(\frac{k+1}{k+2}\right)\frac{1}{2\as^{k+1}\bar c_{(k)}}\left\{
\begin{array}{cc}
1	&	\mbox{ if }	|\Delta_k|\leq \as^{k+1}\bar c_{(k)}\\[10pt]
\frac{1}{(|\Delta_k|/(\as^{k+1}\bar c_{(k)}))^{k+2}}	&	\mbox{ if }	|\Delta_k|>\as^{k+1}\bar c_{(k)}
\end{array}
\right. \, .
\ee
This result depends on the entire set of the calculated coefficients via the parameter $\bar c_{(k)} = \max(|c_0|,\dots,|c_k|)$. 

The knowledge of $f(\Delta_k |c_0,\dots,c_k)$ allows one to calculate the smallest $p\%$-credible interval for  $\Delta_k$. It turns out to be centred at zero, and hence we denote it by
 $[-d_k^{(p)},d_k^{(p)}]$. It is defined implicitly by 
\be
\label{eq:pconf}
p\% = \int_{-d_k^{(p)}}^{d_k^{(p)}}  f(\Delta_k |c_0,\dots,c_k) \ud\Delta_k 
\ee
and one finds, using the analytical approximation in eq.~(\ref{eq:DeltaKnowCkExpression}) (see appendix~\ref{app:pCLint})
\be\label{eq:pCLint}
d_k^{(p)}= \left\{
\begin{array}{cc}
\as^{k+1}\bar c_{(k)}\frac{k+2}{k+1}p\%                   &	\mbox{ if }	 p\% \le \frac{k+1}{k+2}\\[10pt]
\as^{k+1}\bar c_{(k)}\left[(k+2)(1-p\%)\right]^{-1/(k+1)} &	\mbox{ if }	 p\% > \frac{k+1}{k+2} 
\end{array}
\right.
\ee
where, of course, $p\% \equiv p/100$ and $p$ is a number between 0 and 100.

The result for $f(\Delta_k|c_0,\dots,c_k)$ in eq.~(\ref{eq:DeltaKnowCkExpression}) can be generalised to any choice of $f(c_n|\bar c)$ and $f_\epsilon(\bar c)$, i.e. beyond the choices of eqs.~(\ref{eq:cbar}) and (\ref{eq:epsdep}). Using the derivation given in appendix \ref{app:DeltaKnowCkExpression} we obtain, still within the approximation of eq.~(\ref{eq:approxDelta}), 
\be
\label{eq:DeltaKnowCkGeneral}
f_\epsilon(\Delta_k|c_0,\dots,c_k) = \frac{1}{f_\epsilon(c_0,\dots,c_k)} \frac{1}{\as^{k+1}} \int f(c_0|\bar c)...f(c_k|\bar c)\ f(c_{k+1}=\frac{\Delta_k}{\as^{k+1}}|\bar c)\ f_\epsilon(\bar c)\ \ud \bar c\, ,
\ee
where we have now explicitly allowed for the possibility of expressing intermediate quantities as a function of $\epsilon$ (eq.~(\ref{eq:DeltaKnowCkExact}) was instead  already written in the $\epsilon\to 0$ limit).
This expression will be used for the numerical evaluations of densities and credible intervals proposed in the Mathematica package available from the authors.

\begin{figure}[tp]
\begin{center}
\begin{tabular}{ccc}
\includegraphics[height=3.2cm]{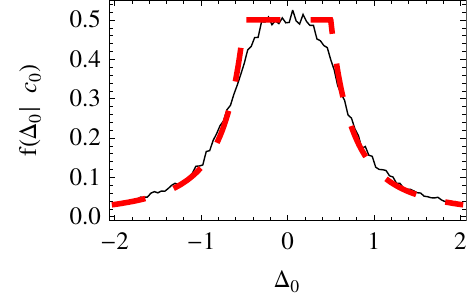}&
\includegraphics[height=3.2cm]{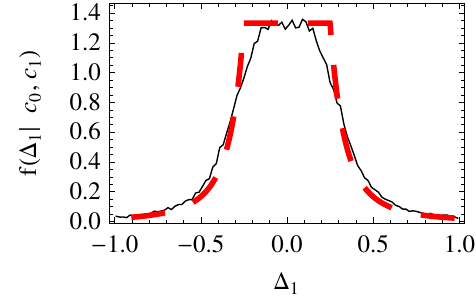}&
\includegraphics[height=3.2cm]{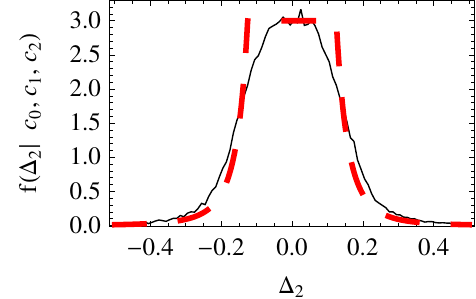}\\
\includegraphics[height=3.2cm]{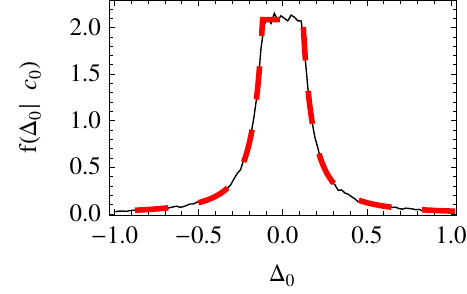}&
\includegraphics[height=3.2cm]{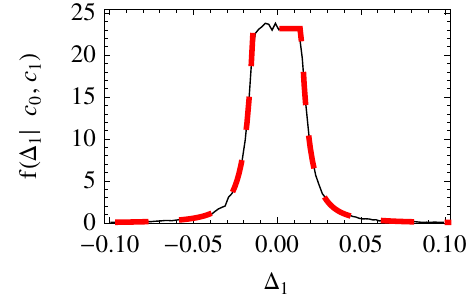}&
\includegraphics[height=3.2cm]{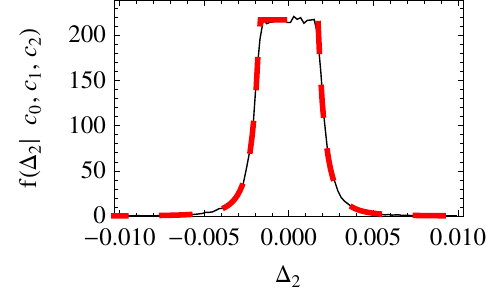}
\end{tabular}
\caption{\label{fig:DeltaApprox}Numerical estimates of the exact densities $f(\Delta_k|c_0,\dots,c_k)$ (continuous curves) and their analytical approximations in eq.~(\protect\ref{eq:DeltaKnowCkExpression}) (dashed curves) in the case $\bar c_{(k)}=1$ for $k=0$ (left), $k=1$ (middle), and $k=2$ (right), for $\as = 0.5$ (top row) and $\as = 0.12$ (bottom row). 
These numerical estimates are computed by integrating over the distributions for  $10$ unknown coefficients, the results being stable when using more. Using values of $\as$ of the order of  0.2 or 0.3 does not degrade significantly the quality of the approximation seen here in the $\as = 0.12$ case.
}
\end{center}
\end{figure}

Since the results in eqs.~(\ref{eq:DeltaKnowCkExpression}) and (\ref{eq:DeltaKnowCkGeneral})  were obtained using the approximation in  eq.~(\ref{eq:approxDelta}), we now wish to check it by comparing them to numerical estimates of the exact density~(\ref{eq:DeltaKnowCkExact}). In order to do so
we  perform a numerical integration of eq.~(\ref{eq:DeltaKnowCkExact}), rewritten in the form
\be
f(\Delta_k|c_0,\dots,c_k)=\int \left[\delta(\Delta_k-\sum_{n=k+1}^{\infty}\as^n c_n)\right]
\left[\prod_{n=k+1}^\infty f(c_n|\bar c)\right] f(\bar c| c_0,\dots,c_k)\ud\bar c\,
\ud c_{k+1} \ud c_{k+2} \dots
\ee
where $f(\bar c| c_0,\dots,c_k)$ is given in eq.~(\ref{eq:cbarKnowCk}) and the $f(c_n|\bar c)$ in eq.~(\ref{eq:cbar}).
Figure \ref{fig:DeltaApprox} shows the numerical results for $k=0$, $1$ and $2$ and the corresponding analytical approximation for $f(\Delta_k|c_0,\dots,c_k)$ in eq.~(\ref{eq:DeltaKnowCkExpression}). We can see that the agreement is extremely good, especially when small (realistic) value of $\as$ are used. We will therefore rely on the approximation of equation~(\ref{eq:approxDelta}) for our predictions of densities for $\Delta_k$ in the rest of this paper.


\section{Comparison with the conventional method}
\label{sect:compVarMu}

In deriving the density for $\Delta_k$ in the previous section we made no reference to the scale variation $\delta_k$ of the partial sum $\sigma_k(Q,\mu)$ which is usually employed in the conventional uncertainty estimate $[\sigma_k^-,\sigma_k^+]$ of section \ref{sect:conv}. In order to assess the compatibility of the two methods, we now wish to study the relation between the density for $\Delta_k$ and an interval of the kind $[\sigma_k^-,\sigma_k^+]$.

Given a specific series and a set of coefficients $(c_0,\dots,c_k)$ we wish to evaluate
\be 
\label{eq:defConfDeltakAncienne}
\mathbb{C}(\Delta_k\in[\Delta_k^-,\Delta_k^+]|c_0,\dots,c_k)
	=\int_{\Delta_k^-}^{\Delta_k^+} f(\Delta_k|c_0,\dots,c_k)\ \ud\Delta_k
\ee
and, for definiteness, we now take $[\sigma_k^-,\sigma_k^+]$ as the interval given by eq.~(\ref{eq:sigmak1}), so that we can set
\ba
&&\Delta_k^-=\min(\sigma_k(Q,Q/2),\sigma_k(Q,2Q))-\sigma_k = \sigma_k^--\sigma_k \\
&&\Delta_k^+=\max(\sigma_k(Q,Q/2),\sigma_k(Q,2Q))-\sigma_k = \sigma_k^+-\sigma_k
\ea
Since the  shape of $\sigma_k(Q,\mu)$, and therefore the values of $\Delta_k^-$ and $\Delta_k^+$, depend on all the values of the calculated coefficients $(c_0,\dots,c_k)$,  
while the density function $f(\Delta_k|c_0,\dots,c_k)$ depends only on their maximum $\bar c_{(k)}$ (see eq.~(\ref{eq:DeltaKnowCkExpression})), it is  important to make sure that different sets of coefficients, sharing the same $\bar c$, do not typically lead to broadly different  estimates for the degree of belief  $\mathbb{C}(\Delta_k\in[\Delta_k^-,\Delta_k^+]|c_0,\dots,c_k)$ in eq. (\ref{eq:defConfDeltakAncienne}).

\begin{figure}[tp]
\begin{center}
\begin{tabular}{ccc}
\includegraphics[height=3.3cm]{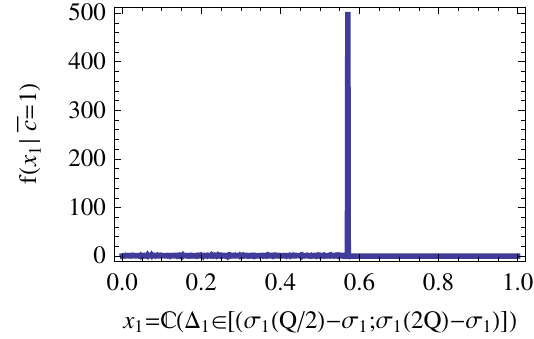}&
\includegraphics[height=3.3cm]{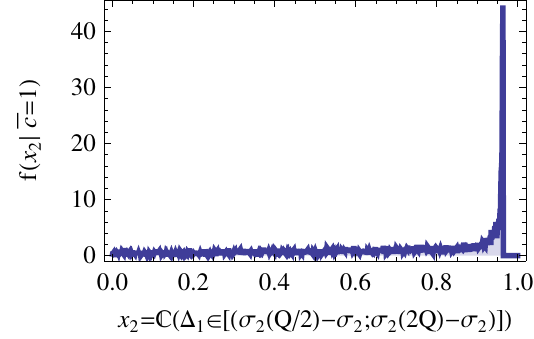}&
\includegraphics[height=3.3cm]{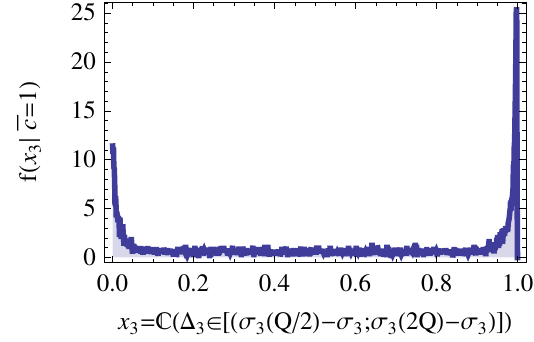}\\
\end{tabular}
\caption{Numerical estimates of the exact densities $f(x_k|\bar c)$, where $x_k(c_0,\dots,c_k)=\mathbb{C}(\Delta_k\in[\Delta_k^-,\Delta_k^+]|c_0,\dots,c_k)$ is the degree of belief of the scale variation  interval, for $\bar c=1$,  for $k=1$ (left) and $k=2$ (middle) and $k=3$ (right). Each plot is obtained using  $N=10^4$ samples.}
\label{fig:xApprox}
\end{center}
\end{figure}

To test this in practice, we evaluate the integral (\ref{eq:defConfDeltakAncienne})
for many different configurations of the coefficients $(c_0,\dots,c_k)$, in the case $\bar c = 1$. Figure~\ref{fig:xApprox} shows the distribution of the values of the degrees of belief that are obtained for the three perturbative orders $k=1,2,3$.
The typical degree of belief $\mathbb{C}(\Delta_k\in[\Delta_k^-,\Delta_k^+]|c_0,\dots,c_k)$ predicted by the model can be seen to be largely unaffected by the precise values of the coefficients, its distribution taking an almost Dirac-delta shape at the values 0.57, 0.96 and 0.99 for $k=1,2,3$ respectively. The peak at $x_3\simeq 0$ in the rightmost plot can be understood as an artifact due to configurations where $\sigma_3(Q,2Q)$ and $\sigma_3(Q,Q/2)$ are accidentally close to each other, resulting in a vanishing $[\Delta_k^-,\Delta_k^+]$ interval. It can be made to disappear by modifying the choice of the interval and using instead $[\Delta_k^-,\Delta_k^+]=[\min\{\sigma_k(Q,\mu)\}-\sigma_k,\max\{\sigma_k(Q,\mu)\}-\sigma_k]$, i.e. corresponding to eq.~(\ref{eq:sigmak2}) rather than eq.~(\ref{eq:sigmak1}).

The numerical results of figure~\ref{fig:xApprox} can be understood through the following analytical approximations.
First, we modify slightly the interval $[\Delta_k^-,\Delta_k^+]$ considered above. We make it symmetric around $\sigma_k$ and of width $\delta_k/2$, where $\delta_k$ is given in eq.~(\ref{eq:approxdelta1}),
and we consider
\be
\mathbb{C}(\Delta_k\in[\Delta_k^-,\Delta_k^+]|c_0,\dots,c_k)
	\simeq \mathbb{C}(\Delta_k\in[-\frac{\delta_k}{2},\frac{\delta_k}{2}]|c_0,\dots,c_k)\label{eq:approxdeltabar}
\ee
Using eq.~(\ref{eq:DeltaKnowCkExpression}), we get (see appendix~\ref{app:conflev})
\be
\mathbb{C}(\Delta_k\in[-\frac{ \delta_k}{2},\frac{\delta_k}{2}]|c_0,\dots,c_k) =
\left\{
\begin{array}{cc}	
1-\frac{1}{k+2}\left[\frac{2}{3k\beta_0}\frac{\bar c_{(k)}}{|c_k|}\right]^{k+1}
& \mbox{ if }\frac{\delta_k}{2} \ge \as^{k+1}\bar c_{(k)} \Leftrightarrow |c_k| \ge \frac{2}{3k\beta_0} \bar c_{(k)} 
\label{eq:deltaKConf}
\\[10pt]
\frac{k+1}{k+2} \frac{3 k\beta_0}{2}\frac{|c_k|}{\bar c_{(k)}} 
& \mbox{ if }\frac{\delta_k}{2} < \as^{k+1}\bar c_{(k)} \Leftrightarrow |c_k| < \frac{2}{3k\beta_0} \bar c_{(k)} 
\end{array}
\right.
\ee
This result is fully independent of the coefficients in the approximation (or in the case) where
$c_k = \bar c_{(k)}$. It predicts the Dirac distribution-like shape observed in figure~\ref{fig:xApprox} and the variation of its position with the value of $k$. For $k=1$, $k=2$ and $k=3$, it gives for the degree of belief~(\ref{eq:deltaKConf}) the values  $61\%$ (using the lower expression in eq.~(\ref{eq:deltaKConf})), $96\%$ and $99.6\%$ (using the upper expression in eq.~(\ref{eq:deltaKConf})) respectively, using $\beta_0 = 0.61$ and $c_k = \bar c_{(k)}$. These values are in good agreement with those obtained from the numerical estimates of the exact densities in figure~\ref{fig:xApprox}.
The $k$-dependence of the result in eq.~(\ref{eq:deltaKConf}) shows that the degree of belief of the interval  $[\sigma_k - \delta_k/2,\sigma_k + \delta_k/2]$ is not a constant,  but depends instead on the perturbative order at which we are working. When calculating higher orders in a perturbative series not only the size of the conventional residual uncertainty decreases, but also its degree of belief, as evaluated by our new method, increases. Note also that this method of evaluating the degree of belief of an uncertainty interval avoids one specific shortfall of the conventional method, namely that its  estimate of the theoretical uncertainty may become unreasonably small if the last calculated coefficient happens by accident to be much smaller than the others or if accidental cancellations take place.

It would be tempting to consider eq.~(\ref{eq:deltaKConf}) as the main results from our Bayesian model, allowing one to associate a degree of belief to the uncertainty bands given by the conventional method. The simplicity of these equations, and their numerical values tantalisingly (though entirely accidentally) close to the confidence levels of Gaussian sigmas, make them apparently good candidates for such an identification. However, it is important to bear in mind that these equations depend on the  choice made for the density function in eq.~(\ref{eq:cbar}), as well as on the various approximations made in deriving them. As such, they cannot be considered as strictly valid in general, although they offer a very useful first approximation when trying to gauge the degree of belief of a conventional uncertainty band generated by scale variations.

In practice, one would like to be able to abandon the scale variations method altogether, and  determine the degree of belief of any interval of his choosing. In general we will therefore not use eq.~(\ref{eq:deltaKConf}), but rather 
estimate any desired $p\%$-credible interval numerically using the density function~(\ref{eq:DeltaKnowCkGeneral}), without any reference to the conventional method.

\section{Series starting at non-zero order $\as^l$}
\label{sect:non-zero-start}
Oftentimes, one may wish to consider a perturbative series starting at a non-zero order in $\as$,
\be
\sigma = \sum_{n=l}^\infty c_n\as^n \, .
\ee
When this is the case, only $k+1-l$ coefficients (rather than $k+1$) are calculated when the series is known up to perturbative order $k$. The results of our model given in the previous section should then be modified as follows.

Eqs.~(\ref{eq:cbarKnowCk1}) and (\ref{eq:cbarKnowCk}) become
\be
\label{eq:cbarKnowCk1-mod}
f(\bar c|c_l,\dots,c_k)=n_c\ \frac{(\max(|c_l|,\dots,|c_k|))^{n_c}}{\bar c^{n_c+1}}\ \chi_{\bar c>(\max(|c_l|,\dots,|c_k|))}
=n_c\ \frac{\bar c_{(k)}^{n_c}}{\bar c^{n_c+1}}\ \chi_{\bar c > \bar c_{(k)}}
\ee
where we have introduced the {\sl number} of known coefficients, 
\be
n_c \equiv k+1-l \, .
\ee
Note also that 
$\bar c_{(k)}$ should now formally be defined as $\max(|c_l|,\dots,|c_k|)$. We have not changed its notation so as not to proliferate the number of different symbols.

Eqs.~(\ref{eq:cnKnowCk1}) and (\ref{eq:fcn}) are similarly modified to
\ba
\label{eq:cnKnowCk1-mod}
f(c_n|c_l,\dots,c_k) &=&
\frac{1}{2}\frac{n_c}{n_c+1}\ \frac{(\max(|c_l|,\dots,|c_k|))^{n_c}}{(\max(|c_n|,|c_l|,\dots,|c_k|))^{n_c+1}}
\nonumber\\
&=&
\left(\frac{n_c}{n_c+1}\right)\frac{1}{2\bar c_{(k)}}\left\{
\begin{array}{cc}
1	&	\mbox{ if }	|c_n|\leq \bar c_{(k)}\\
\frac{1}{(|c_n|/\bar c_{(k)})^{n_c+1}}	&	\mbox{ if }	|c_n|>\bar c_{(k)}
\end{array}
\right. \, .
\ea

Eq.~(\ref{eq:DeltaKnowCkExpression}) becomes
\be
\label{eq:DeltaKnowCkExpression-mod}
f(\Delta_k|c_l,\dots,c_k)\simeq\left(\frac{n_c}{n_c+1}\right)\frac{1}{2\as^{k+1}\bar c_{(k)}}\left\{
\begin{array}{cc}
1	&	\mbox{ if }	|\Delta_k|\leq \as^{k+1}\bar c_{(k)}\\[10pt]
\frac{1}{(|\Delta_k|/(\as^{k+1}\bar c_{(k)}))^{n_c+1}}	&	\mbox{ if }	|\Delta_k|>\as^{k+1}\bar c_{(k)}
\end{array}
\right. 
\ee
and from this one derives the result corresponding to eq.~(\ref{eq:pCLint}) for the width of the smallest $p$\%-credibility interval:
\be
\label{eq:pCLint-mod}
d_k^{(p)}= \left\{
\begin{array}{cc}
\as^{k+1}\bar c_{(k)}\frac{n_c+1}{n_c}p\%                   &	\mbox{ if }	 p\% \le \frac{n_c}{n_c+1}\\[10pt]
\as^{k+1}\bar c_{(k)}\left[(n_c+1)(1-p\%)\right]^{-1/n_c} &	\mbox{ if }	 p\% > \frac{n_c}{n_c+1} 
\end{array}
\right. \, .
\ee

Finally, using the result in eq.~(\ref{eq:approxdelta1}), $\delta_k \simeq 3 k\beta_0 \as^{k+1}|c_k|$, which is unmodified by the fact that the series starts now at order $l$, one finds that the degree of belief associated to the interval given by the conventional scale-variation method, already given in eq.~(\ref{eq:deltaKConf}) for the $l=0$ case, is modified as
\be
\mathbb{C}(\Delta_k\in[-\frac{ \delta_k}{2},\frac{\delta_k}{2}]|c_l,\dots,c_k) =
\left\{
\begin{array}{cc}	
1-\frac{1}{n_c+1}\left[\frac{2}{3k\beta_0}\frac{\bar c_{(k)}}{|c_k|}\right]^{n_c}
& \mbox{ if }\frac{\delta_k}{2} \ge \as^{k+1}\bar c_{(k)} \Leftrightarrow |c_k| \ge \frac{2}{3k\beta_0} \bar c_{(k)} 
\label{eq:deltaKConf-mod}
\\[10pt]
\frac{n_c}{n_c+1} \frac{3 k\beta_0}{2}\frac{|c_k|}{\bar c_{(k)}} 
& \mbox{ if }\frac{\delta_k}{2} < \as^{k+1}\bar c_{(k)} \Leftrightarrow |c_k| < \frac{2}{3k\beta_0} \bar c_{(k)} 
\end{array}
\right.
\ee

For a process starting at order $\as$ (i.e. $l=1$) this equation predicts a degree of belief of 46\% at LO ($k=1$), using the lower expression in eq.~(\ref{eq:deltaKConf-mod}),  90\%  at NLO ($k=2$) and 98.8\% at NNLO ($k=3$), using in both cases the lower expression in eq.~(\ref{eq:deltaKConf-mod}) and $c_k = \bar c_{(k)}$. For a process starting at order $\as^2$ (i.e. $l=2$) one predicts instead a degree of belief of 73\% at LO ($k=2$), 96\%  at NLO ($k=3$) and 99.5\% at NNLO ($k=4$). In this case the upper expression in eq.~(\ref{eq:deltaKConf-mod}) always applies.


\section{A realistic application: $e^+e^-\to$ hadrons}
\label{sect:results}

The total cross section  $\sigma(e^+e^-\to\gamma\to$~hadrons) is one of the best known observables in perturbative QCD, its coefficients being known exactly up to order $\as^3$, and even $c_4$ being known approximately. This process is therefore an ideal place where to test the behaviour of our  Bayesian model, and compare it to the results of the conventional uncertainty estimate.

We write this cross section as
\be
\sigma_4(Q) = \sigma_0(Q) (1 + \sum_{n=1}^4 c_n \as^n(Q))
\ee
and, for $n_f = 5$ massless flavours, we have (the values are reviewed for instance in~\cite{pdg})
\be
c_1 = 0.31831,\qquad
c_2 = 0.142785,\qquad
c_3 = -0.412969,\qquad
c_4 \simeq -0.821356
\ee
These coefficients leads to following partial sums\footnote{Note that in \cite{pdg} $\sigma_{QCD}$ is denoted by $\delta_{QCD}$ instead. We have modified the notation to avoid confusion with our own definition for $\delta_k$, which represents an uncertainty interval rather than a value of the truncated series.}
for $\sigma_{QCD,k}(Q) \equiv \sigma_k(Q)/\sigma_0(Q) -1$ and $\as(Q) = 0.118$:
\be
\sigma_{QCD,1} = 0.0375606,\qquad
\sigma_{QCD,2} = 0.0395487,\qquad
\sigma_{QCD,3} = 0.0388702
\ee
where we have dropped for convenience the argument $Q$.
One can now apply the conventional uncertainty estimate method of section~\ref{sect:conv}. Using for definiteness the convention in eq.~(\ref{eq:sigmak2}), one finds
\ba
&&[\sigma_{QCD,1}^-,\sigma_{QCD,1}^+] = [0.03401,~0.04197] \label{eq:conv-intervals1}\\
&&[\sigma_{QCD,2}^-,\sigma_{QCD,2}^+] = [0.03871,~0.03980] \label{eq:conv-intervals2}\\
&&[\sigma_{QCD,3}^-,\sigma_{QCD,3}^+] = [0.03855,~0.03893] \label{eq:conv-intervals3}
\ea
One can compare these ``uncertainty intervals'' with the successive perturbative results given above, and see that indeed order by order the higher-order result is inside the interval\footnote{These uncertainty intervals have been evaluated by using in all cases an evolution equation for $\as$ up to $\beta_2$, i.e. what is needed for $\sigma_{QCD,3}$. One may have used lower-order accuracies when dealing with  $\sigma_{QCD,1}$ or  $\sigma_{QCD,2}$, but we have explicitly checked that this changes at most the last significant figure in the numbers given above.}.

This is as far as the conventional uncertainty estimate can go. At this point one can use our model to do one of two things (or both): either one calculates the degree of belief of the intervals given above, or one finds the intervals corresponding to given values of degree of belief, for instance 68.3\%, 95.5\% and 99.7\%.

\begin{figure}[t]
\begin{center}
\includegraphics[width=11cm]{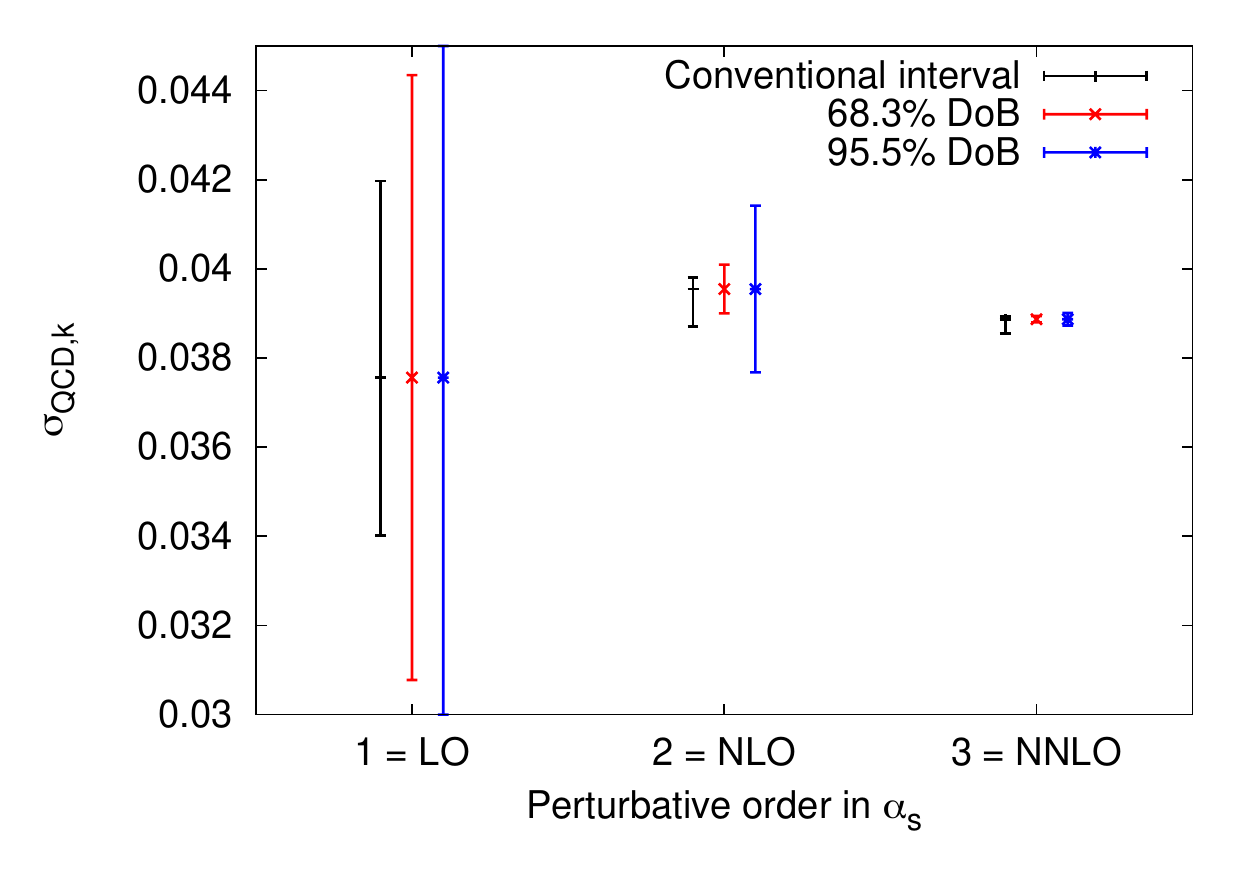}
\caption{\label{fig:intervals} Comparison of the uncertainty intervals for the $e^+e^-\to$~hadrons process, as  given by the conventional method of scale variations (first interval on the left of each group) and by our model (the latter for two different values of degree of belief, 68.3\% and 95.5\%, respectively middle and right of each group), for $\as = 0.118$. We have used the definition $\sigma_{QCD,k}(Q) \equiv \sigma_k(Q)/\sigma_0(Q) -1$.}
\end{center}
\end{figure}

For the first option, we find
\ba
&&{\mathbb C}(\sigma_{QCD} \in [\sigma_{QCD,1}^-,\sigma_{QCD,1}^+]|c_1) = 45.8\%\\
&&{\mathbb C}(\sigma_{QCD} \in [\sigma_{QCD,2}^-,\sigma_{QCD,2}^+]|c_1,c_2) = 58.4\%\\ 
&&{\mathbb C}(\sigma_{QCD} \in [\sigma_{QCD,3}^-,\sigma_{QCD,3}^+]|c_1,c_2,c_3) = 77.2\% 
\ea
These values have been obtained via numerical integration of the density $f(\Delta_k|c_1,\dots,c_k)$ in eq.~(\ref{eq:DeltaKnowCkExpression-mod}). One can compare them to the values given by the analytical approximations in eq.~(\ref{eq:deltaKConf-mod}) for $l=1$ and $k=1,2,3$ respectively, i.e. $n_c = k+1-l =1,2,3$. One obtains 45.8\%, 54.8\% and 98.8\% respectively.
The first two values (both obtained with the lower expression in eq.~(\ref{eq:deltaKConf-mod})) can be seen to be in good agreement with the exact results. The big discrepancy for the third one can be explained with the fact that the actual interval $[\sigma_{QCD,3}^-,\sigma_{QCD,3}^+]$ happens to be quite asymmetric with respect to the central value $\sigma_{QCD,3}$, whereas in the approximation (\ref{eq:deltaKConf-mod}) the interval $[-\delta_k/2,\delta_k/2]$ is symmetric. This serves as a remainder that one should always resort to the full numerical evaluation of the degree of belief whenever accurate results are sought.

For the second option, we find, using the notation ${\mathbb C}_k^{(p)}$ to denote the minimal $p\%$-credible interval of the form $[\sigma_k-d_k^{(p)},\sigma_k+d_k^{(p)}]$, where $d_k^{(p)}$ is defined implicitly in eq.~(\ref{eq:pconf}) (see also appendix~\ref{app:pCLint}),
\ba
&&{\mathbb C}^{(68.3)}_1 = [0.0307747,~0.0443465]\\
&&{\mathbb C}^{(68.3)}_2 = [0.0390012,~0.0400962]\\
&&{\mathbb C}^{(68.3)}_3 = [0.0387973,~0.0389431]
\ea
and
\ba
&&{\mathbb C}^{(95.5)}_1 = [-0.0023475,~0.0774687]\\
&&{\mathbb C}^{(95.5)}_2 = [0.0376789,~0.0414185]\\
&&{\mathbb C}^{(95.5)}_3 = [0.0387297,~0.0390107]
\ea
These intervals\footnote{These numerical values have been calculated discarding the coefficient of $\as^0$, on the ground that it controls an exclusively electroweak process, and therefore it should not have a say on the size of the coefficients of a perturbative expansion in $\as$. They differ very slightly (at the level of the third/fourth significant figure) from those that can be obtained using eq.~(\ref{eq:pCLint-mod}), since they have been calculated by integrating numerically the credibility distributions. Note that in the definition of the model (and therefore in the numerical evaluation of the credible intervals) one may employ for $f(c_n|\bar c)$ a form which differs from the uniform step function given in eq.~(\ref{eq:cbar}), for instance a Gaussian distribution of mean 0 and standard deviation $\bar c$. We have checked that in this case the intervals are not modified in a significant way, showing that the results of our model are robust with respect to reasonable variations of this initial hypothesis.}
can then be compared to those returned by the conventional method in eqs.~(\ref{eq:conv-intervals1},\ref{eq:conv-intervals2},\ref{eq:conv-intervals3}). This comparison is given in graphical form in figure~\ref{fig:intervals}. One can see how the 68.3\%-credible  intervals are not too dissimilar from those predicted by the conventional method of scale variations. It is worth noting how the former tend to become smaller than the latter as the perturbative order increases, pointing to a potential overestimate of the theoretical uncertainty by the conventional method at higher orders.


\section{Discussion about the hypotheses of the model}
\label{sect:modifModel}

Our model was built making the choices in eqs.~(\ref{eq:cbar}) and~(\ref{eq:epsdep}) for the  densities $f(c_n|\bar c)$ and $f_\epsilon(\bar c)$. We made there the choice of using a flat prior for $\ln \bar c$ (rather than $\bar c$ itself) in eq.~(\ref{eq:epsdep}),
and for $c_n$ instead in eq.~(\ref{eq:cbar}). We discuss below the reasoning behind these choices.

\subsection{Choice of the density function $f(c_n|\bar c)$}

The choice of exactly what variable to use to express a prior density, e.g. the logarithm of a parameter rather than the parameter itself, is related to the assumed nature of said parameter. Suppose we used $\ln c_n$   instead of $c_n$ in eq.~(\ref{eq:cbar}) and defined a uniform  density 
\be\label{eq:cnLn}
f(\ln c_n|\bar c,h)=\frac{1}{2 h}\chi_{\ln \bar c - h\leq\ln c_n\leq \ln \bar c+ h}
\ee
where $h$ is an arbitrary parameter.
This means to consider it as likely to find $c_n$ between $\bar c/\exp(h)$ and $\bar c$ as between $\bar c$ and $\bar c \exp(h)$. One can debate whether this behaviour  is more or less appropriate than the one used in  eq.~(\ref{eq:cbar}) where $c_n$ is equally likely to lie between $-\bar c$ and zero as between zero and $\bar c$.
However, the main drawback of eq.~(\ref{eq:cnLn}) is that it requires the introduction of a new, a priori unknown, parameter $h$ which controls the spread of the coefficients. At least three perturbative coefficients would then need to be known before the model can estimate a credibility interval.

We have therefore concluded that the hypothesis in eq.~(\ref{eq:cbar}) not only already describes sufficiently well the observed typical relations between perturbative coefficients, but 
also provides the simplest model (simplicity being a strong guiding principle of our model, as we wish to be able to control well the information we introduce into it).

\subsection{Choice of the density function $f(\ln\bar c)$}

The value of the sum of a perturbative series depends on the value of $\bar c$. Choosing a  density for $\bar c$ which is uniform in $\bar c$ itself rather than in its logarithm amounts to trying to predict the precise value of such a series rather than just its order of magnitude. We find the former too strong a constraint, and prefer therefore to limit ourselves to the second choice. On a technical side, we also find that when using a prior uniform in $\bar c$ one then needs at least two calculated coefficients in order to have a non-null  density on its theoretical uncertainty, whereas in the $\ln \bar c$ case one coefficient is already sufficient to give an indication about the order of magnitude of the higher order coefficients and therefore about the remainder of the series.

Note also that it is sufficient to use in eq.~(\ref{eq:epsdep}) a  density  $f_\epsilon(\ln\bar c)$ that is uniform in $\ln\bar c$ only in the $\epsilon\to 0$ limit. For finite values of $\epsilon$ this requirement is not necessary.

\subsection{Choice of the expansion parameter}

Another modification of the model would be to use an expansion parameter that differs from $\as$, so that
\be
\sigma = \sum c_n \as^n= \sum (\lambda^n c_n)\left(\frac{\as}{\lambda}\right)^n
\ee
This corresponds to a redefinition of the coefficients $c_n$ into
\be
c'_n=\lambda^n c_n
\ee
and the density function in eq.~(\ref{eq:cbar}) would now be defined by
\be
f(c'_n|\bar c')=\frac{1}{2\bar c}\chi_{|c'_n|\leq\bar c'}
\ee
where $\bar c'$ is a parameter that applies to the new set of coefficients $c_n'$.

This choice would not modify the expressions for the densities over the unknown coefficients $f(c'_n|c'_0,\dots,c'_k)$, but only the one over the residual sum $f(\Delta_k|c'_0,\dots,c'_k)$ since the approximation~(\ref{eq:approxDelta}) would now read
\be
\Delta_k\simeq c'_{k+1}\left(\frac{\as}{\lambda}\right)^{k+1}
\ee
so that eq.(\ref{eq:DeltaKnowCkExpression}) is replaced by
\be\label{eq:DeltaKnowCkExpressionPrime}
f(\Delta_k|c_0',\dots,c_k')\simeq\left(\frac{k+1}{k+2}\right)\frac{1}{2(\as/\lambda)^{k+1}\bar c'_{(k)}}\left\{
\begin{array}{cc}
1	&	\mbox{ if }	|\Delta_k|\leq (\as/\lambda)^{k+1}\bar c'_{(k)}\\
\frac{1}{(|\Delta_k|/((\as/\lambda)^{k+1}\bar c'_{(k)}))^{k+2}}	&	\mbox{ if }	|\Delta_k|>(\as/\lambda)^{k+1}\bar c'_{(k)}
\end{array}
\right. \, ,
\ee
where now  $\bar c'_{(k)}$ is of course the maximum of the new known coefficients $c_n'$.

Different values for $\lambda$ relate to different speeds of convergence of the series, either quicker ($\lambda>1$) or slower ($\as<\lambda<1$). Of course one must be careful not to end up with an expansion parameter which is too large, because this will eventually invalidate the use of the approximation in eq.~(\ref{eq:approxDelta}).


\section{Partially known higher orders}
\label{sect:limModel}

The model we have considered so far assumes perfect knowledge of some coefficients, up to order $k$, and total ignorance of those of higher order. In practice, it is often possible to know {\sl part} of a higher order coefficient, typically calculated within some approximation or obtained as an expansion of an all-order resummation. It is yet straightforward to extend the model to account for such  cases. 

Two new building blocks are required to adapt the model. First of all, if $\tilde c_{k+1}$ is an approximation of $c_{k+1}$ it should not provide more information than the true value $c_{k+1}$ itself. If the real value $c_{k+1}$ is known, knowledge of the approximate value $\tilde c_{k+1}$ must not change anything: for a set of coefficients $\{c_i\}$ it must hold
\begin{align}
\label{eq:partiallyknowncoeffs}
f(\{c_i\}|c_{k+1},\tilde{c}_{k+1})
	&=f(\{c_i\}|c_{k+1})\, ,\\
f(\bar c|c_{k+1},\tilde c_{k+1})
	&=f(\bar c|c_{k+1})\, ,\\
f(\bar c,\{c_i\}|c_{k+1},\tilde c_{k+1})
	&=f(\bar c,\{c_i\}|c_{k+1})\, .
\end{align}

Secondly, one must decide how reliable a given approximation $\tilde c_{k+1}$ of $c_{k+1}$ is. We must introduce a  density function $f(\tilde c_{k+1}|c_{k+1})$ for the  value $\tilde c_{k+1}$, given the true $c_{k+1}$. The choice of this  density will depend on the way $\tilde c_{k+1}$ was obtained. One possible parametrisation is for instance the log-normal density
\be\label{eq:likely}
f(\tilde{c}_{k+1}|c_{k+1})= \frac{1}{|\tilde{c}_{k+1}|}\frac{1}{\sqrt{2\pi}\ln f}\exp\left(-\frac{(\ln(\tilde{c}_{k+1}/c_{k+1}))^2}{2(\ln f)^2}\right)\, .
\ee
for some chosen value of the parameter $f$. It more or less corresponds to $\tilde c_{k+1}$ estimating $c_{k+1}$ up to a factor of order $f$. 

The  densities on the true value of the coefficient $c_{k+1}$ which is known only approximately, and on the completely unknown coefficients $c_n$  can then be written, up to normalisation factors collectively denoted by $\mathcal{N}$, as (see appendix~\ref{app:approxcoeff})
\ba
&&f(c_{k+1}|c_0,\dots,c_k,\tilde c_{k+1})
	=\mathcal{N} f(c_{k+1}|c_0,\dots,c_k)f(\tilde c_{k+1}|c_{k+1})\\
&&f(c_n|c_0,\dots,c_k,\tilde c_{k+1})
	=\mathcal{N}\int f(c_n,c_{k+1}|c_0,\dots,c_k) f(\tilde c_{k+1}|c_{k+1})\ \ud c_{k+1}
\label{eq:approxcoeff}
\ea
More generally, for arbitrary sets of known coefficients $C_K=\{c_i\}_{i\in [0,k]}$, approximations $\tilde C_A=\{\tilde c_i\}_{i\in A}$ of coefficients $C_A=\{c_i\}_{i\in A}$ and  totally unknown coefficients $C_N=\{c_i\}_{i\in N}$ we can write
\be
f(C_N,C_A|C_K,\tilde C_A)
	=\mathcal{N}\   f(C_N,C_A|C_K)\   f(\tilde C_A|C_A)
\label{eq:densCoefPartGene}
\ee
To get the  density for the unknown coefficients, one then just integrates over $C_A$. To get the  density over $\Delta_k$ one replaces eq.~(\ref{eq:densCoefPartGene}) in the definition~(\ref{eq:DeltaKnowCkExact}).

Let us study for instance the case of one known coefficient $c_0$ and one approximation $\tilde c_1$. We want to obtain the density $f(\Delta_0|c_0,\tilde c_1)$. Depending on how much we trust the approximation, the uncertainty over $c_1$ or $c_2$ will predominate. In general, we need to keep track of both these coefficients in the expression for $\Delta_0$. We do not use the approximation~(\ref{eq:approxDelta}) but rather the more accurate one
\be
\Delta_0\simeq c_1\as+ c_2 \as^2 \, .
\ee
The result for the  density is then
\be\label{eq:densDeltaKnowC0C1Approx}
f(\frac{\Delta_0}{\as}|c_0,\tilde c_1)\simeq f(c_1+ c_2 \as=\frac{\Delta_0}{\as}|c_0,\tilde c_1)
\ee
The expression for $f(c_1+ c_2 \as=x|c_0,\tilde c_1)$ can be obtained as 
\begin{align}
f(c_1+\as c_2=x|c_0,\tilde c_1)
	&=\int \delta(x-(c_1+\as c_2))f(c_2,c_1|c_0,\tilde c_1)\ \ud c_2\ \ud c_1\nonumber\\
	&=\int f(c_2,c_1=x-\as c_2|c_0,\tilde c_1)\ \ud c_2\nonumber\\
	&=\int f(c_2,c_1(x,c_2)|c_0)f(\tilde c_1|c_1(x,c_2))\ \ud c_2
\end{align}
where we defined $c_1(x,c_2)\equiv x-\as c_2$ to simplify the expression and we used equation~(\ref{eq:densCoefPartGene}). The result for  $f(c_2,c_1(x,c_2)|c_0)$ is obtained in a similar way to $f(c_n|c_0,\dots,c_k)$ in~(\ref{eq:fcn}):
\be
f(c_2,c_1(x,c_2)|c_0)=\lim_{\epsilon\rightarrow 0}\frac{f_\epsilon(c_2,c_1(x,c_2),c_0)}{f_\epsilon(c_0)}
\ee
Using eq.~(\ref{eq:c0ckeps}) for $k=0$ and $k=2$ we find
\be
\label{eq:toto}
f(c_1+\as c_2=x|c_0,\tilde c_1)=\mathcal{N}\int \left(\frac{1}{\bar c_{(2)}(x)}\right)^3\ f(\tilde c_1|c_1(x,c_2))\ \ud c_2
\ee
where we have defined $\bar c_{(2)}(x) \equiv \max(c_0,c_1(x,c_2),c_2)$.

\begin{figure}[tp]
\begin{center}
\begin{tabular}{cccc}
\includegraphics[width=3.7cm]{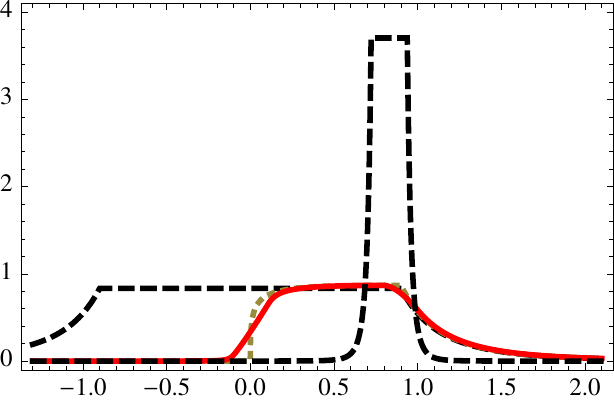}&
\includegraphics[width=3.7cm]{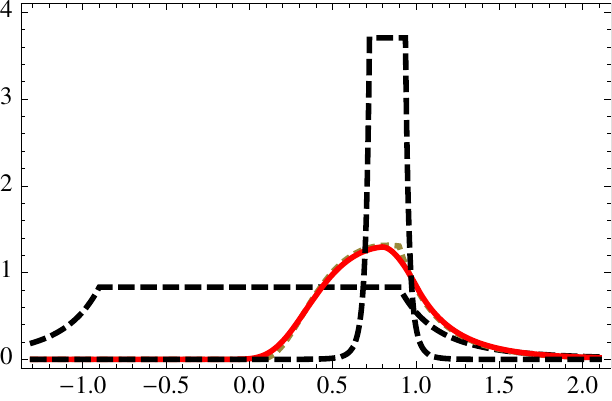}&
\includegraphics[width=3.7cm]{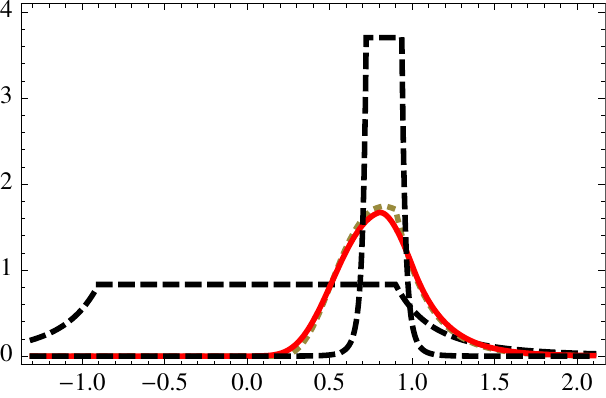}&
\includegraphics[width=3.7cm]{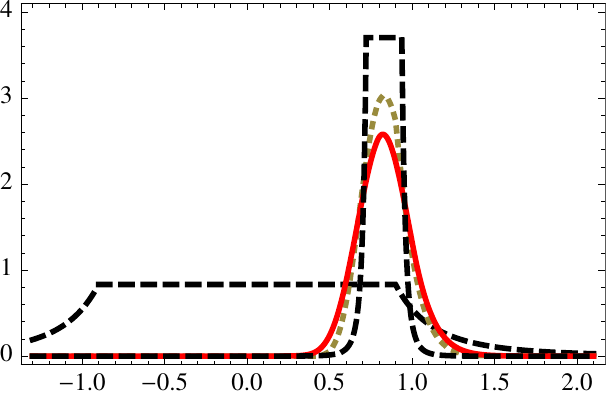}\\
\includegraphics[width=3.7cm]{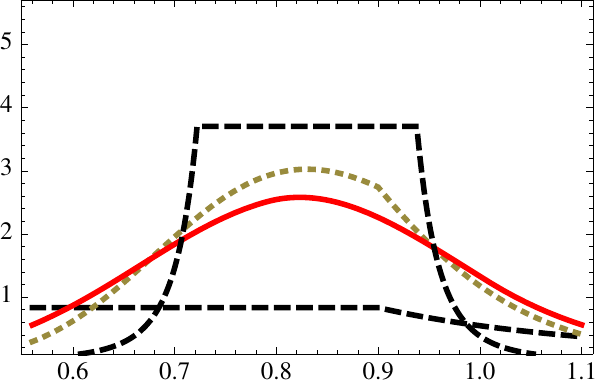}&
\includegraphics[width=3.7cm]{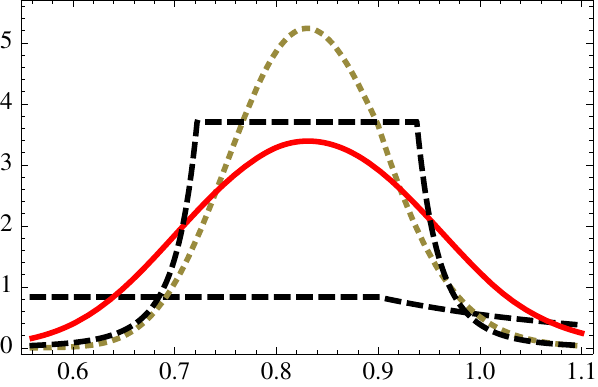}&
\includegraphics[width=3.7cm]{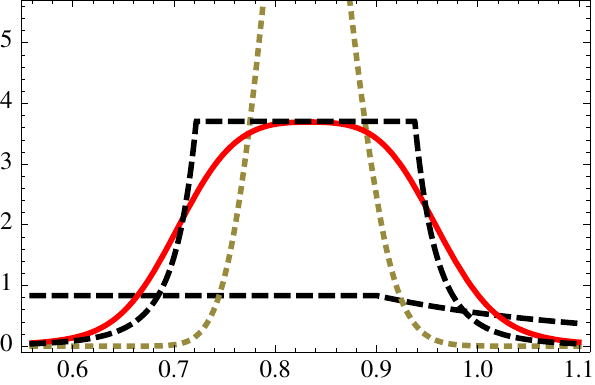}&
\includegraphics[width=3.7cm]{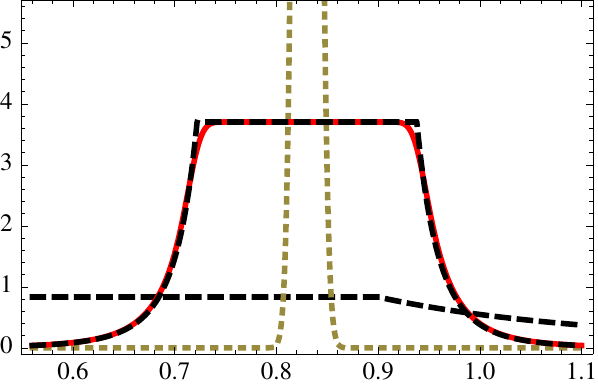}
\end{tabular}
\end{center}
\caption{Density $f(c_1+\as c_2|c_0,\tilde c_1)\simeq f(\frac{\Delta_0}{\as}|c_0,\tilde c_1)$ (solid curve) for $c_0=0.9$, $\tilde{c}_1=0.83$, $\as=0.12$ and a log-normal density $f(\tilde c_1|c_1)$ of width parameter $\ln f$ with $f=50$ ; $2$ ; $1.5$ ; $1.2$, from left to right in the top row, and $f=1.2$ ; $1.1$ ; $1.05$ ; $1.01$, from left to right in the bottom row. The two dashed curves represent $2f(c_1|c_0)\simeq2f(\frac{\Delta_0}{\as}|c_0)$ (flattest dashed curve) and $f(c_1+\as c_2|c_0,c_1=\tilde c_1)\simeq f(\frac{\Delta_0}{\as}|c_0,c_1=\tilde c_1)$ (most peaked dashed curve). They correspond to the limits when $\tilde c_1$ is not trusted at all or when it is completely trusted. The last, dotted curve represents $f(c_1|c_0,\tilde c_1)$. It coincides with $f(\frac{\Delta_0}{\as}|c_0,\tilde c_1)$ until the uncertainty over $c_1$ is the limiting one.}
\label{fig:extApprox}
\end{figure}

In order to see how this works in practice, we choose $c_0=0.9$ and $\tilde c_1=0.83$ and we plot eq.~(\ref{eq:toto}) as a function of $x=\Delta_0/\as$ for various values of $f$.
Figure~\ref{fig:extApprox} shows how the density is modified when $\tilde c_1$ is more and more trusted ($f$ is smaller and smaller). Consider at first the top left plot. When the approximation is not much trusted and $f$ is large ($f = 50$ in this case), the only useful information that we can get from $\tilde c_1$ is the sign of $c_1$. The total uncertainty over $\Delta_0/\as\simeq c_1+\as c_2$ is largely dominated by the uncertainty over $c_1$. Its  density coincides with two times the density $f(c_1|c_0)$ for positive values of $x$ and with $f(c_1|c_0,\tilde c_1)$. The differences, originating from the uncertainty over $c_2$ or from the small information provided by the particular value of $\tilde c_1$, are almost negligible (except around zero).

As $\tilde c_1$ gets more trusted, the uncertainty decreases and the degree of belief for $\Delta_0/\as$ to have a value around $\tilde c_1$ increases (top middle plots). The uncertainty over $c_1$ is still the limiting one but the information provided by $\tilde c_1$ is not negligible anymore. The full density still coincides with $f(c_1|c_0,\tilde c_1)$ but is starting to be different from $f(c_1|c_0)$.

Then comes a limit where the uncertainty over $c_1$ is of the same order as the one over $\as c_2$ (top-right and bottom-left plots. They are the same plot, but shown for different scales on the x-axis). The full  density over $c_1+\as c_2$ now differs from $f(c_1|c_0,\tilde c_1)$ and its width is given both by this density and by the one over $f(c_2|c_0,c_1=\tilde c_1)$. 

As $\tilde c_1$ is considered to be a better and better approximation, the uncertainty over $c_2$ prevails. The difference between $c_1$ and $\tilde c_1$ is now negligible and the total  density approaches the shape of $f(c_1+\as c_2|c_0,c_1=\tilde c_1)$ (bottom middle and right plots).


\section{Conclusions and outlook}

In this paper we have introduced a Bayesian model which allows one to characterise in terms of intervals of a given degree of belief 
(or degree of belief of a given interval) 
the residual theoretical uncertainty of a perturbative calculation. 
Our aim is to put on more solid ground the estimate of the uncertainty of a known result, not to improve in any way the calculation itself. This we try to achieve by formalising hypotheses on the behaviour of the coefficients of perturbative series, and then by deriving from these hypotheses the degree of belief values in a rigorous way.

We have chosen to try to translate as closely as possible into our model the assumption which is implicitly made when employing the conventional method (scale variations) for estimating the uncertainty, namely that successive coefficients of a perturbative series tend to have similar size. One may or may not believe this hypothesis to be well grounded, and our choice is not necessarily true or even just the best possible one. 
However, what matters, and what this paper wants to provide, is not so much which hypotheses are made, 
but rather the formalism that allows one to derive from them a proper characterisation of the residual theoretical uncertainty: our framework can then be considered as a box into which to input one's favourite hypothesis about the behaviour of a series, and from which to extract the appropriate degree of belief values.

We have found that, under the quite general assumption mentioned above and within the Bayesian framework
formalised in section~\ref{sect:model}, the   $p$\%-credible interval  $[\sigma_k - d^{(p)}_k,\sigma_k + d^{(p)}_k]$ of a series calculated up to order $k$,
\be
\sigma_k = c_l\as^l  + \dots + c_k\as^k \, ,
\ee
with $l \ge 0$ and $k \ge l$, is given by
\be
d_k^{(p)}= \left\{
\begin{array}{cc}
\as^{k+1}\max\{|c_l|,\dots, |c_k|\}\frac{n_c+1}{n_c}p\%                   &	\mbox{ if }	 p\% \le \frac{n_c}{n_c+1}\\[10pt]
\as^{k+1} \max\{|c_l|,\dots, |c_k|\} \left[(n_c+1)(1-p\%)\right]^{-1/n_c} &	\mbox{ if }	 p\% > \frac{n_c}{n_c+1} 
\end{array}
\right. \, ,
\ee
with $n_c = k+1-l$ the number of known perturbative coefficients. The full credibility  distribution can also be obtained, and is given in section~\ref{sect:deltak}.

In the calculation of QCD corrections to a simple process like $e^+e^-\to$~hadrons we see
that the intervals given by the conventional renormalisation scale variation are not too
dissimilar from the 68.3\%-credible intervals given by our Bayesian model. These
findings, detailed in section~\ref{sect:results} and shown in graphical form in
figure~\ref{fig:intervals}, are perhaps not surprising: the conventional method itself has
been built and refined over the years into a form that often returns results compatible
with the calculation of successive perturbative orders and with intuitive expectations, 
and the same hypothesis that it makes implicitly we have made explicitly. Nevertheless,
within our method one can now state a precise interval and in addition a
detailed degree of belief for it (and possibly bet on it). A Mathematica package implementing the results of this paper is available from the authors.

Obviously this is not the final word in terms of a rigorous characterisation of 
theoretical uncertainties. We have chosen a very simple process, and we have
found a nice self-consistent picture. However, much more work remains to be done
in order to extend the method to more complex processes. For one thing, one may wish to accommodate also the presence of a
factorisation scale, and therefore additional ingredients like parton
distribution or fragmentation functions. Secondly, our Bayesian model as it is
now formulated inevitably fails when the behaviour of a physical process is
known not to be self-similar to all orders: an obvious example is a process for
which a new production channel opens at some perturbative order, or for which a particular kinematical configuration is selected.  In such cases, an extension of our hypotheses is obviously called for. We are looking
forward to exploring these new avenues in the future.

\vspace{15pt}
\noindent
{\bf Acknowledgments.} We wish to thank Giulio D'Agostini and Gavin Salam for useful comments on the manuscript. This work was supported in part by grants ANR-09-BLAN-0060 and PITN-GA-2010-264564.

\appendix

\section{Partial cross section, coefficients and renormalisation scale}
\label{app:coefs}

Let
\be
\sigma(Q)=\sum_{i=0}^{\infty}c_i(Q,\mu)\as^i(\mu)
\ee
be the total sum of a perturbative series of expansion parameter $\as$, which evolves according to
\be
\frac{\ud\as}{\ud\ln\mu^2} = \beta(\as) = -\as^2\sum_{j=0}^\infty \beta_j \as^j \; .
\ee
where $\beta(\as)$ is the beta function. $\mu$ is an arbitrary renormalisation scale, so that it holds
\begin{eqnarray}
\frac{\ud \sigma}{\ud \ln\mu^2}=0
&=&\sum_{i=0}^\infty \left(\frac{\ud c_i}{\ud \ln\mu^2}\as^i-ic_i\as^{i-1}\sum_{j=0}^{\infty}\beta_j\as^{j+2}\right)\label{eq:derivSigma}\nonumber\\
&=&\sum_{i=0}^\infty\as^i\left[\frac{\ud c_i}{\ud \ln\mu^2}-\sum_{j=0}^{i-1}j\beta_{i-1-j}c_j\right]\label{eq:derivsum}
\end{eqnarray}
so that
\be\label{eq:derivcoeff}
\frac{\ud c_i}{\ud \ln\mu^2}=\sum_{j=0}^{i-1}j\beta_{i-1-j}c_j\qquad\forall i\geq0
\ee
This equation already tells us that the $\mu$-dependence of the coefficient $c_i$ is controlled by the lower-order coefficients $c_j$, $j \le i-1$. Moreover, it shows that $c_0$ and $c_1$ are independent of $\mu$. It is then straightforward to conclude that $c_i$ is a polynomial of degree less than or equal to $i-1$ in $\ln\frac{\mu^2}{Q^2}$:
\be
c_i(Q,\mu)=\sum_{l=0}^{i-1}c_{i,l}\left(\ln\frac{\mu^2}{Q^2}\right)^l\qquad\forall i\geq 1
\ee
This is true for $i=1$ since the derivative of $c_1$ with respect to $\mu$ is zero. Assuming that it is true up to some $i$, eq.~(\ref{eq:derivcoeff}) shows that $\frac{\ud c_{i+1}}{\ud \ln\mu^2}$ is a polynomial in $\ln\frac{\mu^2}{Q^2}$ of degree less than or equal to $i-1$, which makes $c_{i+1}$ itself a polynomial of degree less than or equal to $i$. 

Rewriting eq.~(\ref{eq:derivcoeff}) order by order in $\ln\frac{\mu^2}{Q^2}$, we can also obtain a recurrence relation giving the values of all $c_{i,l}$ in terms  of the $c_{j,0} = c_j(Q,Q)$, $j<i$, and the $\beta_j$:
\be
\label{eq:derivPartialSum}
c_{i,l}=\frac{1}{l}\sum_{j=0}^{i-1}j\beta_{i-1-j}c_{j,l-1}
\ee
Hence, given the calculated values for $c_i(Q,Q)$ one can easily reconstruct the full renormalisation scale dependence of the coefficients and of the partial sums $\sigma_k = \sum_{i=0}^k c_i\as^i$. Once again, note that only the coefficient $c_i(Q,Q) = c_{i,0}$ needs to be explicitly computed at each order.

Finally, we give the expression of the derivative of a partial sum $\sigma_k$ with respect to $\ln\frac{\mu^2}{Q^2}$. From eqs.~(\ref{eq:derivsum}) and (\ref{eq:derivcoeff}) we get
\be
\frac{\ud \sigma_k}{\ud \ln\mu^2}=\sum_{i=k+1}^\infty\as^i\sum_{j=0}^k j\beta_{i-1-j} c_j
\ee
showing that, as expected, the residual scale dependence of $\sigma_k$ is of higher order $\as^{k+1}$.
If we now consider that $\as \ll 1$, observe that the known coefficients $\beta_i$ are such that $\beta_i \lesssim \beta_0$ (see table \ref{tab:beta}), and assume all the $|c_i|$ are of the same order, we can approximate this result as
\be
\frac{\ud \sigma_k}{\ud \ln\mu^2}  \simeq\as^{k+1}k\beta_{0} c_k \, .
\ee

\begin{table}[tp]
\begin{center}
$$
\begin{array}{|c||c|c|c|c|}
\hline
k		&	i=1	&	i=2	&	i=3	&	i=4	\\\hline\hline
1		&	0.61	&	-	&	-	&	-	\\\hline
2		&	0.24	&	1.22	&	-	&	-	\\\hline
3		&	0.07	&	0.49	&	1.83	&	-	\\\hline
4		&	0.19	&	0.15	&	0.73	&	2.44	\\\hline
\end{array}
$$
\caption{First values of $i\beta_{k-i}$, calculated with $n_f = 5$. The first column gives the first four coefficients of the beta function.}
\label{tab:beta}
\end{center}
\end{table}

\section{Derivations of density distributions and uncertainty intervals}
\label{app:derivDens}

\subsection{Derivation of $f(c_n|c_0,\dots,c_k)$ in eq.~(\ref{eq:cnKnowCk1})}
\label{app:cnKnowCk1}

We wish to derive derive eq.~(\ref{eq:cnKnowCk1}). We first compute the density $f_\epsilon(c_n|c_0,\dots,c_k)$ for $\epsilon\neq 0$. Using the definition of conditional density, we have
\be
f_\epsilon(c_n|c_0,\dots,c_k)=\frac{f_\epsilon(c_0,\dots,c_k,c_n)}{f_\epsilon(c_0,\dots,c_k)}\qquad\forall n>k
\ee
Both densities $f_\epsilon(c_0,\dots,c_k)$ and $f_\epsilon(c_0,\dots,c_k,c_n)$ are obtained from the independence hypothesis (\ref{eq:modIndep}) and the expressions of the densities of the model (\ref{eq:cbar}) and (\ref{eq:epsdep}). We have then
\begin{eqnarray}
f_\epsilon(c_0,\dots,c_k)
	&=&\int f_\epsilon(c_0,\dots,c_k,\bar c) \ \ud \bar c\nonumber\\
	&=&\int f_\epsilon(c_0,\dots,c_k|\bar c) f_\epsilon(\bar c) \ \ud \bar c\nonumber\\
	&=&\int \left[\prod_{i=0}^k f(c_i|\bar c)\right] f_\epsilon(\bar c)\ \ud \bar c\nonumber\\
	&=&\int \left[\prod_{i=0}^k \frac{1}{2\bar c}\ \chi_{|c_i|\leq\bar c}\right] \left[\frac{1}{2|\ln\epsilon|}\frac{1}{\bar c}\chi_{\epsilon\leq\bar c\leq1/\epsilon}\right]\ud\bar c\nonumber\\	&=&\frac{1}{2^{k+2}}\frac{1}{|\ln\epsilon|}\int_{\max(|c_0|,\dots,|c_k|,\epsilon)}^{1/\epsilon}\frac{1}{\bar c^{k+2}}\ \ud \bar c \, .
\end{eqnarray}
A similar calculation gives for $f_\epsilon(c_0,\dots,c_k,c_n)$:
\be
f_\epsilon(c_0,\dots,c_k,c_n)	=\frac{1}{2^{k+3}}\frac{1}{|\ln\epsilon|}\int_{\max(|c_0|,\dots,|c_k|,|c_n|,\epsilon)}^{1/\epsilon}\frac{1}{\bar c^{k+3}}\ \ud \bar c \, .
\ee
We can then obtain the expression for the density $f_\epsilon(c_n|c_0,\dots,c_k)$
\be
f_\epsilon(c_n|c_0,\dots,c_k)=\frac{1}{2}\frac{-\frac{1}{k+2}[\bar c^{-(k+2)}]_{\max(|c_0|,\dots,|c_k|,|c_n|,\epsilon)}^{1/\epsilon}}{-\frac{1}{k+1}[\bar c^{-(k+1)}]_{\max(|c_0|,\dots,|c_k|,\epsilon)}^{1/\epsilon}}
\ee
and, going to the limit $\epsilon\rightarrow0$, 
\be
f(c_n|c_0,\dots,c_k)=\frac{1}{2}\frac{k+1}{k+2}\frac{\max(|c_0|,\dots,|c_k|)^{k+1}}{\max(|c_0|,\dots,|c_k|,|c_n|)^{k+2}} \, .
\ee
Note that in this equation the value $k+1$ represents the total {\sl number} of known perturbative coefficients $c_0,\dots,c_k$ used to estimate $c_n$ with $n > k$, rather than simply one unit above the last calculated perturbative order $k$. Similarly, $k+2$ is this total number plus one. If a series starts at a non-zero order $\as^l$, its last known perturbative order $k$ plus one will not give anymore the number of known coefficients. We detail in section~\ref{sect:non-zero-start} the modifications to be made to account for such a case.

\subsection{Derivation of $f(\bar c|c_0,\dots,c_k)$ in eq.~(\ref{eq:cbarKnowCk1})}
\label{app:cbarKnowCk1}

The result~(\ref{eq:cbarKnowCk1}) is obtained in a way similar to the one discussed in appendix \ref{app:cnKnowCk1}. We find
\begin{eqnarray}
f_\epsilon(\bar c|c_0,\dots,c_k)
	&=&\frac{f_\epsilon(\bar c,c_0,\dots,c_k)}{f_\epsilon(c_0,\dots,c_k)}\nonumber\\
	&=&\frac{
	\phantom{{}\int{}}
	f_\epsilon(\bar c,c_0,\dots,c_k)
	\phantom{\ \ud \bar c}
	}
	{
	\int{}
	f_\epsilon(\bar c,c_0,\dots,c_k)
	\ \ud \bar c
	}\nonumber\\
	&=&\frac{
	\phantom{{}\int{}}
	f_\epsilon(c_0,\dots,c_k|\bar c)f_\epsilon(\bar c)
	\phantom{\ \ud \bar c}
	}
	{
	\int 
	f_\epsilon(c_0,\dots,c_k|\bar c)f_\epsilon(\bar c)
	\ \ud \bar c
	}\nonumber\\
	&=&\frac{
	\frac{1}{2^{k+2}}
	\frac{1}{|\ln\epsilon|}
	\phantom{{}\int{}}
	\frac{1}{\bar c^{k+2}}
	\chi_{\max(|c_0|,\dots,|c_k|,\epsilon)\leq\bar c\leq1/\epsilon}
	\phantom{\ \ud \bar c}
	}
	{
	\frac{1}{2^{k+2}}
	\frac{1}{|\ln\epsilon|}
	\int
	\frac{1}{\bar c^{k+2}}
	\chi_{\max(|c_0|,\dots,|c_k|,\epsilon)\leq\bar c\leq1/\epsilon}
	\ \ud \bar c 
	} \, .
\end{eqnarray}
Taking the limit $\epsilon\rightarrow0$ and using the notation $\bar c_{(k)}=\max(|c_0|,\dots,|c_k|)$ we get
\be
f(\bar c|c_0,\dots,c_k)=(k+1)\frac{\bar c_{(k)}^{k+1}}{\bar c^{k+2}}\chi_{\bar c\geq\bar c_{(k)}} \, .
\ee

\subsection{Derivation of the smallest $p$\%-credible interval in eq.~(\ref{eq:pCLint})}
\label{app:pCLint}

The density $f(\Delta_k|c_0,\dots,c_k)$ in eq.~(\ref{eq:DeltaKnowCkExpression}) is symmetric for negative and positive $\Delta_k$, and decreases monotonically  from $\Delta_k=0$ to infinity (see figure~\ref{fig:DeltaApprox}). The smallest  interval of fixed $p\%$ degree of belief, which we denote by $[-d_k^{(p)},d_k^{(p)}]$,
will then also be symmetric. 

Two cases apply. With $p$ sufficiently large, this interval will extend beyond the $[-\as^{k+1}\bar c_{(k)},\as^{k+1}\bar c_{(k)}]$ range, so that the density's expression in eq.~(\ref{eq:DeltaKnowCkExpression}) can be simplified as
\be
f(\Delta_k|c_0,\dots,c_k)=\left(\frac{k+1}{k+2}\right)\frac{1}{2\as^{k+1}\bar c_{(k)}}
\frac{1}{(|\Delta_k|/(\as^{k+1}\bar c_{(k)}))^{k+2}}\qquad \mathrm{for}\quad \Delta_k\not\in[-d_k^{(p)},d_k^{(p)}]
\ee
Noting that the degree of belief outside of $[-d_k^{(p)},d_k^{(p)}]$ is $1-p\%$ and that the interval and the density are symmetric, we have
\begin{align}
\frac{1-p\%}{2}
	&=\int_{d_k^{(p)}}^\infty  f(\Delta_k |c_0,\dots,c_k) \ud\Delta_k\nonumber\\
	&=\int_{d_k^{(p)}}^\infty\left(\frac{k+1}{k+2}\right)\frac{1}{2\as^{k+1}\bar c_{(k)}}\frac{1}{(|\Delta_k|/(\as^{k+1}\bar c_{(k)}))^{k+2}}\ud\Delta_k\nonumber\\
	&=\frac{1}{2}\left(\frac{k+1}{k+2}\right)(\as^{k+1}\bar c_{(k)})^{k+1}\int_{d_k^{(p)}}^\infty\frac{\ud\Delta_k}{\Delta_k^{k+2}}\nonumber\\
	&=\frac{1}{2}\left(\frac{k+1}{k+2}\right)(\as^{k+1}\bar c_{(k)})^{k+1}\frac{1}{k+1}\left(\frac{1}{d_k^{(p)}}\right)^{k+1}\label{eq:confExp}
\end{align}
From this we obtain
\be
\label{eq:case1}
d_k^{(p)}=\as^{k+1}\bar c_{(k)}[(k+2)(1-p\%)]^{-1/(k+1)}
\ee

If, on the other hand, the interval is smaller than the $[-\as^{k+1}\bar c_{(k)},\as^{k+1}\bar c_{(k)}]$ range, it is the upper expression in eq.~(\ref{eq:DeltaKnowCkExpression}) that enters the calculation of the degree of belief. One finds
\be
p\% = \int_{-d_k^{(p)}}^{d_k^{(p)}}  f(\Delta_k |c_0,\dots,c_k) \ud\Delta_k = 
 \int_{-d_k^{(p)}}^{d_k^{(p)}} \left(\frac{k+1}{k+2}\right)\frac{1}{2\as^{k+1}\bar c_{(k)}} \ud\Delta_k = 
 d_k^{(p)}\left(\frac{k+1}{k+2}\right)\frac{1}{\as^{k+1}\bar c_{(k)}}
\ee
so that
\be
\label{eq:case2}
d_k^{(p)} = \as^{k+1}\bar c_{(k)}\frac{k+2}{k+1}p\%
\ee

One can see that the first case, leading to eq.~(\ref{eq:case1}), applies for $p\% > \frac{k+1}{k+2}$, whereas the second one, leading to eq.~(\ref{eq:case2}), holds for $p\% \le \frac{k+1}{k+2}$.

\subsection{Derivation of the approximate $f(\Delta_k|c_0,\dots,c_k)$ in eq.~(\ref{eq:DeltaKnowCkGeneral})}
\label{app:DeltaKnowCkExpression}

From eq.~(\ref{eq:DeltaKnowCkExact}) and making the approximation $\Delta_k\simeq\as^{k+1}c_{k+1}$ we get
\begin{align}
f(\Delta_k|c_0,\dots,c_k)
	&=\int \left[\delta(\Delta_k-\sum_{n=k+1}^{\infty}c_n\as^n)\right] \prod_{n=k+1}^{\infty}f(c_{n}|\bar c)\ f(\bar c|c_0,\dots,c_k)\ \ud \bar c\ \ud c_{k+1}\ud c_{k+2}\dots\nonumber\\
	&\simeq\int \left[\delta(\Delta_k-c_{k+1}\as^{k+1})\right] \prod_{n=k+1}^{\infty}f(c_{n}|\bar c)\ f(\bar c|c_0,\dots,c_k)\ \ud \bar c\ \ud c_{k+1}\ud c_{k+2}\dots\nonumber\\
	&=\int \left[\delta(\Delta_k-c_{k+1}\as^{k+1})\right] f(c_{k+1}|\bar c)\ f(\bar c|c_0,\dots,c_k)\ \ud \bar c\ \ud c_{k+1}\nonumber\\
	&=\frac{1}{\as^{k+1}}\int f(c_{k+1}=\frac{\Delta_k}{\as^{k+1}}|\bar c)\ f(\bar c|c_0,\dots,c_k)\ \ud \bar c
\end{align}
We can reinstate explicitly the $\epsilon$-dependence in the equation above, and
rewrite the density $f(\bar c|c_0,\dots,c_k)$ in terms of the elementary densities of the model, so that 
the resulting expression can be used with any density distributions. We obtain
\begin{align}
f_\epsilon(\Delta_k|c_0,\dots,c_k)
	&=\frac{1}{\as^{k+1}}\int f(c_{k+1}=\frac{\Delta_k}{\as^{k+1}}|\bar c)\ \frac{f_\epsilon(c_0,\dots,c_k|\bar c)f_\epsilon(\bar c)}{f_\epsilon(c_0,\dots,c_k)}\ \ud \bar c\nonumber\\
	&=\frac{1}{f_\epsilon(c_0,\dots,c_k)}\frac{1}{\as^{k+1}}\int f(c_{k+1}=\frac{\Delta_k}{\as^{k+1}}|\bar c)\ f(c_0|\bar c)\dots f(c_k|\bar c)f_\epsilon(\bar c)\ \ud \bar c
\end{align}
Under this form, the evaluation of $f_\epsilon(\Delta_k|c_0,\dots,c_k)$ can be performed numerically with $\epsilon\neq 0$.

\subsection{Derivation of the degree of belief of the scale variation bands in eq.~(\ref{eq:deltaKConf})}
\label{app:conflev}

The result in eq.~(\ref{eq:deltaKConf}) for the degree of belief of an interval $[-\frac{\delta_k}{2},\frac{ \delta_k}{2}]$ can be easily obtained by recalling the derivation in appendix \ref{app:pCLint} of eq.~(\ref{eq:pCLint}) and inverting the final result, i.e. expressing the degree of belief as a function of the interval width rather than viceversa. One easily obtains
\be
\mathbb{C}(\Delta_k\in[-\frac{ \delta_k}{2},\frac{\delta_k}{2}]|c_0,\dots,c_k) =
\left\{
\begin{array}{cc}	
1-\frac{1}{k+2}\left[\frac{\as^{k+1}\bar c_{(k)}}{\delta_k/2}\right]^{k+1}
& \mbox{ if }\frac{\delta_k}{2} \ge \as^{k+1}\bar c_{(k)} 
\\[10pt]
\frac{k+1}{k+2} \frac{\delta_k/2}{\as^{k+1}\bar c_{(k)}}
& \mbox{ if }\frac{\delta_k}{2} < \as^{k+1}\bar c_{(k)} 
\end{array}
\right.
\ee
and, using the result in eq.~(\ref{eq:approxdelta1}),
\be
\delta_k \simeq  3 k\beta_0 \as^{k+1}|c_k| \, ,
\ee
we get
\be
\mathbb{C}(\Delta_k\in[-\frac{ \delta_k}{2};\frac{\delta_k}{2}]|c_0,\dots,c_k) =
\left\{
\begin{array}{cc}	
1-\frac{1}{k+2}\left[\frac{2}{3k\beta_0}\frac{\bar c_{(k)}}{|c_k|}\right]^{k+1}
& \mbox{ if }\frac{\delta_k}{2} \ge \as^{k+1}\bar c_{(k)} \Leftrightarrow |c_k| \ge \frac{2}{3k\beta_0} \bar c_{(k)} 
\\[10pt]
\frac{k+1}{k+2} \frac{3 k\beta_0}{2}\frac{|c_k|}{\bar c_{(k)}} 
& \mbox{ if }\frac{\delta_k}{2} < \as^{k+1}\bar c_{(k)} \Leftrightarrow |c_k| < \frac{2}{3k\beta_0} \bar c_{(k)} 
\end{array}
\right.
\ee

\subsection{Derivation of $f(c_n|c_0,\dots,c_k,\tilde c_{k+1})$ in eq.~(\ref{eq:approxcoeff})}
\label{app:approxcoeff}

Let us first derive the expression of $f(c_{k+1}|c_0,\dots,c_k,\tilde c_{k+1})$.
\begin{align}
f_\epsilon(c_{k+1}|c_0,\dots,c_k,\tilde c_{k+1})
	&=\frac{
	\phantom{{}\int{}}
	f_\epsilon(c_0,\dots,c_k,c_{k+1},\tilde c_{k+1})
	\phantom{\ \ud c_{k+1}}
	}{
	\int
	f_\epsilon(c_0,\dots,c_k,c_{k+1},\tilde c_{k+1})
	\ \ud c_{k+1}
	}\nonumber\\
	&=\frac{
	\phantom{{}\int{}}
	f_\epsilon(c_0,\dots,c_k|c_{k+1},\tilde c_{k+1})f_\epsilon(c_{k+1},\tilde c_{k+1})
	\phantom{\ \ud c_{k+1}}
	}{
	\int
	f_\epsilon(c_0,\dots,c_k|c_{k+1},\tilde c_{k+1})f_\epsilon(c_{k+1},\tilde c_{k+1})
	\ \ud c_{k+1}
	}
\end{align}
Using the fact that when a coefficient is fully known knowing it approximately adds nothing (see eqs.~(\ref{eq:partiallyknowncoeffs}), we can rewrite this as
\begin{align}	
f_\epsilon(c_{k+1}|c_0,\dots,c_k,\tilde c_{k+1})
	&=\frac{
	\phantom{{}\int{}}
	f_\epsilon(c_0,\dots,c_k|c_{k+1})f_\epsilon(c_{k+1},\tilde c_{k+1})
	\phantom{\ \ud c_{k+1}}
	}{
	\int
	f_\epsilon(c_0,\dots,c_k|c_{k+1})f_\epsilon(c_{k+1},\tilde c_{k+1})
	\ \ud c_{k+1}
	}\nonumber\\
	&=\frac{
	\phantom{{}\int}
	[f_\epsilon(c_0,\dots,c_k,c_{k+1})/f_\epsilon(c_{k+1})][f_\epsilon(\tilde c_{k+1}|c_{k+1})f_\epsilon(c_{k+1})]
	\phantom{\ \ud c_{k+1}}
	}{
	\int
	[f_\epsilon(c_0,\dots,c_k,c_{k+1})/f_\epsilon(c_{k+1})][f_\epsilon(\tilde c_{k+1}|c_{k+1})f_\epsilon(c_{k+1})]
	\ \ud c_{k+1}
	}\nonumber\\
	&=\frac{
	\phantom{{}\int{}}
	f_\epsilon(c_0,\dots,c_k,c_{k+1})f_\epsilon(\tilde c_{k+1}|c_{k+1})
	\phantom{\ \ud c_{k+1}}
	}{
	\int
	f_\epsilon(c_0,\dots,c_k,c_{k+1})f_\epsilon(\tilde c_{k+1}|c_{k+1})
	\ \ud c_{k+1}
	}\nonumber\\
	&=\mathcal{N}_\epsilon f_\epsilon(c_{k+1}|c_0,\dots,c_k)f(\tilde c_{k+1}|c_{k+1})
\end{align}
where with $\mathcal{N}_\epsilon$ we denote the normalisation factor.
In the limit $\epsilon\rightarrow0$ we can therefore write
\be
f(c_{k+1}|c_0,\dots,c_k,\tilde c_{k+1})=\mathcal{N} f(c_{k+1}|c_0,\dots,c_k)f(\tilde c_{k+1}|c_{k+1})
\ee
The density $f(c_n|c_0,\dots,c_k,\tilde c_{k+1})$, for $n>k+1$, is then simply
\begin{align}
f(c_n|c_0,\dots,c_k,\tilde c_{k+1})
	&=\int f(c_n,c_{k+1}|c_0,\dots,c_k,\tilde c_{k+1})\ \ud c_{k+1}\nonumber\\
	&=\int f(c_n|c_0,\dots,c_k,\tilde c_{k+1},c_{k+1}) f(c_{k+1}|c_0,\dots,c_k,\tilde c_{k+1})\ \ud c_{k+1}\nonumber\\
	&=\mathcal{N}\int f(c_n|c_0,\dots,c_k,c_{k+1}) f(c_{k+1}|c_0,\dots,c_k)f(\tilde c_{k+1}|c_{k+1})\ \ud c_{k+1}\nonumber\\
	&=\mathcal{N}\int f(c_n,c_{k+1}|c_0,\dots,c_k)f(\tilde c_{k+1}|c_{k+1})\ \ud c_{k+1}
\end{align}

\end{document}